\documentclass[12pt,epsf,qsymbols]{article}
\pdfoutput=1
\usepackage[utf8]{inputenc}
\usepackage[normalem]{ulem}
\usepackage[english]{babel}
\usepackage{tabularx}
\usepackage{array}
\usepackage{graphics}
\usepackage[pdftex]{graphicx}
\usepackage{psfrag}
\usepackage{epsfig}
\usepackage{subfig}
\usepackage{mathtools}
\usepackage{amssymb}
\usepackage{pifont}
\usepackage{bbold}
\usepackage{setspace}
\usepackage{rotating}
\usepackage{colortbl}
\usepackage{longtable}
\usepackage{slashed}
\usepackage{braket}
\usepackage{lineno}
\makeatletter
\usepackage{textcomp}
\usepackage[usenames,dvipsnames]{xcolor}
\usepackage{relsize}
\usepackage{cite}

\usepackage{float}
\usepackage{multirow}
\usepackage{tikz}
\usetikzlibrary{decorations.pathmorphing,shapes}

\usepackage{lipsum}

\usepackage{bbm}
\usepackage{bm}
\usepackage{enumitem}
\usepackage{amsmath}
\usepackage{verbatim}

\usepackage{hyperref}
\hypersetup{colorlinks,citecolor=nicegreen,linkcolor=niceblue}
\hypersetup{colorlinks=true}

\usepackage{pifont}
\newcommand{\cmark}{\ding{51}}%
\newcommand{\xmark}{\ding{55}}%

\usepackage{calligra}

\usepackage{geometry}

\allowdisplaybreaks

\setlongtables

\newcommand{\bea}{\begin{eqnarray}}
\newcommand{\eea}{\end{eqnarray}}
\newcommand{\beq}{\begin{equation}}
{
\newcommand{\eeq}{\end{equation}}
\newcommand{\ec}{\end{center}}
\newcommand{\bc}{\begin{center}}

\newcommand{\tev}{{\rm TeV}}
\newcommand{\gev}{{\rm GeV}}

\newcommand{\pdir}{p\kern -5.2pt\raise 0.2ex\hbox {/}}

\newcommand{\vdir}{v\kern -5.75pt\raise 0.15ex\hbox {/}}
\newcommand{\kdir}{k\kern -5.75pt\raise 0.15ex\hbox {/}}
\newcommand{\epsdir}{\epsilon\kern -5.0pt\raise 0.15ex\hbox {/}}
\newcommand{\bvdir}{\bar{v}\kern -5.75pt\raise 0.15ex\hbox {/}}
\newcommand{\Ddir}{D\kern -7.75pt\raise 0.20ex\hbox {/}}
\newcommand{\Adir}{A\kern -7.75pt\raise 0.20ex\hbox {/}}
\newcommand{\ldir}{l\kern -5.0pt\raise 0.2ex\hbox{/}}
\newcommand{\varepsdir}{\varepsilon\kern -5.5pt\raise 0.15ex\hbox{/}}

\newcommand{\nn}{\nonumber}

\makeatother

\DeclareMathAlphabet{\mathcalligra}{T1}{calligra}{m}{n}
\DeclareFontShape{T1}{calligra}{m}{n}{<->s*[2.2]callig15}{}

\definecolor{niceblue}{rgb}{0.15,0.15,0.6}
\definecolor{nicegreen}{rgb}{0.1,0.5,0.1}
\definecolor{Red}{rgb}{1.,0.,0.}

\definecolor{Green}{rgb}{0.2,.7,0.2}

\begin{document}
\unitlength = 1mm

\thispagestyle{empty} 
\begin{flushright}
\begin{tabular}{l}
{\tt \footnotesize ZU-TH-52/20}\\{\tt \footnotesize MITP-20-078}\\
\end{tabular}
\end{flushright}
\begin{center}
\vskip 1.8cm\par
{\par\centering \textbf{\LARGE  
\Large \bf New Physics effects in leptonic and semileptonic decays}
\vskip 1.2cm\par
{\scalebox{.85}{\par\centering \large  
\sc Damir Be\v{c}irevi\'c$^{a}$, Florentin Jaffredo$^a$, Ana Pe\~nuelas$^{b}$ and Olcyr~Sumensari$^{c}$}
{\par\centering \vskip 0.7 cm\par}
{\sl 
$^a$~{
IJCLab, Pôle Théorie (Bât.~210), CNRS/IN2P3 et Université Paris-Saclay, \\
91405 Orsay, France} \\[0.5em]
$^b$~PRISMA+ Cluster of Excellence \& Mainz Institute for Theoretical Physics,\\ Johannes Gutenberg University, 55099 Mainz, Germany\\[0.5em]
$^c$~{Physik-Institut, Universit\"at Z\"urich, CH-8057 Z\"urich, Switzerland}
}}

{\vskip 1.65cm\par}}
\end{center}

\vskip 0.85cm
\begin{abstract}

We discuss the  possibilities of extracting the constraints on New Physics by using the current data on 
the leptonic and semileptonic decays of pseudoscalar mesons. 
In doing so we use a general low energy Lagrangian that besides the vector and axial operators also includes the (pseudo-)scalar 
and tensor ones. 
In obtaining constraints on New Physics couplings, we combine the experimental information concerning several decay modes 
with the accurate and precise lattice QCD results for the hadronic matrix elements. 
We propose to study new observables that can be extracted from the angular analysis of the semileptonic decays and 
 discuss their values both in the Standard Model and in some specific scenarios of physics beyond the Standard Model. 

\end{abstract}
\newpage
\setcounter{page}{1}
\setcounter{footnote}{0}
\setcounter{equation}{0}
\noindent

\renewcommand{\thefootnote}{\arabic{footnote}}

\setcounter{footnote}{0}

\tableofcontents

\newpage

\section{Introduction}
\label{sec:introduction}

Leptonic and semileptonic decays of hadrons in the Standard Model (SM) are described by the weak charged currents and as such they are useful for extracting 
the values of the Cabibbo--Kobayashi-Maskawa (CKM) matrix elements. This is done through a comparison of the experimentally established decay rates with the corresponding 
theoretical expressions. The most difficult problem on the theory side is to reliably estimate the central values and uncertainties attributed to the hadronic matrix elements. 
In other words, in order to extract the CKM couplings with a (sub-)percent accuracy the uncertainties related to the evaluation of the effects of non-perturbative QCD need to be kept at a (sub-)percent level too. 

Over the past two decades we witnessed a spectacular progress in taming the hadronic uncertainties by means of numerical simulations of QCD on the lattice (LQCD). In particular, the precision determination of quantities which involve the pseudoscalar mesons (decay constants and form factors) has been radically improved~\cite{Aoki:2019cca}. 
This is the main reason why we will focus our discussion onto the semileptonic decays of one pseudoscalar to another pseudoscalar meson and to the leptonic decays of pseudoscalar mesons. 
Similar semileptonic decays to vector mesons would also be very interesting to consider because they offer a larger set of observables that could be used to probe the effects of 
New Physics (NP)~\cite{Becirevic:2019tpx} but the problem is that (i) most of the vector mesons are broad resonances, and (ii) even in the narrow resonance approximation many more hadronic form factors appear in theoretical expressions, making the whole problem much more difficult to handle on the lattice at the level of precision comparable to that achieved with pseudoscalar mesons only. 
The only exceptions to that pattern are the decays $D_s\to\phi\ell\bar{\nu}$ and $B_c\to J/\psi \ell \bar{\nu}$ which have been studied on the lattice in Ref.~\cite{Donald:2013pea} and \cite{Harrison:2020gvo}, respectively.

In this paper we will therefore use the leptonic and semileptonic decays of pseudoscalar mesons to constrain contributions arising from physics beyond the SM. 
An important ingredient in such an analysis is the CKM matrix, the entries of which are extracted from various flavor observables, including the same leptonic and semileptonic decays that we consider as probes of the NP couplings~\cite{Charles:2015gya,Bona:2006ah}. In order to eliminate this uncertainty in the discussion that follows, we will define suitable observables in which the dependence on the CKM matrix elements cancels out completely. An example of such observables are Lepton Flavor Universality (LFU) ratios, which became popular in recent years owing to 
the discrepancies observed in semileptonic $B$-meson decays~\cite{Bifani:2018zmi}. However, these are not the only theoretically clean observables that are independent on the CKM matrix elements. Another possibility is to consider ratios of leptonic and semileptonic observables, based on the same quark-level transitions, which allow us to probe the NP couplings without requiring specific assumptions on the non-universality of the leptonic couplings. Furthermore, one can exploit the detailed angular analysis of a given semileptonic decay, which provides us with complementary information on physics beyond the Standard Model (BSM).

The remainder of this paper is organized as follows: In Sec.~\ref{sec:eft} we extend the Fermi effective theory to include the most general NP effects. This general effective Lagrangian is then used to compute various semileptonic and leptonic observables in Sec.~\ref{sec:semileptonic} and Sec.~\ref{sec:leptonic}, respectively. In Sec.~\ref{sec:numerics} we discuss the SM predictions for the observables based on $K$, $D_{(s)}$ and $B_{(s)}$ mesons decays. These predictions are then confronted with experimtanl data in Sec.~\ref{sec:numerics-bis} to determine the constraints on the NP couplings and to predict new quantities that can be probed experimentally. Our results are summarized in Sec.~\ref{sec:summary}.

\section{Effective Lagrangian}
\label{sec:eft}

The most general low-energy effective Lagrangian of dimension-six describing the $d_i\to u_j \ell \bar{\nu}$ transition, with $\ell\in\lbrace e,\mu,\tau\rbrace$, is given by
\begin{align}
\label{eq:left}
    \mathcal{L}_\mathrm{eff} &= -2\sqrt{2}G_F V_{ij}\Big{[}(1+g_{V_L}^{ij\,\ell})\,\big{(}\bar{u}_{Li}\gamma_\mu d_{Lj} \big{)}\big{(}\bar{\ell}_L \gamma^\mu\nu_{L}\big{)}+g_{V_R}^{ij\,\ell}\,\big{(}\bar{u}_{Ri}\gamma_\mu d_{Rj} \big{)}\big{(}\bar{\ell}_L \gamma^\mu\nu_{L}\big{)}\\[0.35em]
    &\hspace*{-0.4em}+g_{S_L}^{ij\,\ell}\,\big{(}\bar{u}_{Ri} d_{Lj} \big{)}\big{(}\bar{\ell}_R \nu_{L}\big{)}+g_{S_R}^{ij\,\ell}\,\big{(}\bar{u}_{Li} d_{Rj} \big{)}\big{(}\bar{\ell}_R\nu_{L}\big{)}+g_{T}^{ij\,\ell}\,\big{(}\bar{u}_{Ri}\sigma_{\mu\nu} d_{Lj} \big{)}\big{(}\bar{\ell}_R \sigma^{\mu\nu}\nu_{L}\big{)}\,\Big{]}+\mathrm{h.c.}\,,\nonumber
\end{align}
where $i,j$ denote quark-flavor indices, $V_{ij}$ are the CKM matrix elements and $g_\alpha^{ij\,\ell}$ stand for the effective NP couplings, with $\alpha \in \lbrace V_{L(R)}, S_{L(R)}, T \rbrace$. Neutrinos are assumed to be purely left-handed particles and only lepton flavor conserving transitions are considered. To describe low-energy processes, it is convenient to define effective coefficients with definite parity in the quark current, namely,
\begin{align}
g_{V(A)}^{ij\,\ell} &= g_{V_R}^{ij\,\ell} \pm g_{V_L}^{ij\,\ell}\,,\qquad\qquad g_{S(P)}^{ij\,\ell} = g_{S_R}^{ij\,\ell} \pm g_{S_L}^{ij\,\ell}\,.
\end{align}
\noindent which is useful since the leptonic decays of pseudoscalar mesons will only be sensitive to $g_A^{ij\,\ell}$ and $g_P^{ij\,\ell}$. The remaining effective coefficients, $g_V^{ij\,\ell}$, $g_S^{ij\,\ell}$ and $g_T^{ij\,\ell}$, can be probed by studying the semileptonic processes, $P\to P^\prime \ell \bar{\nu}$, where $P^{(\prime)}$ denote two pseudoscalar mesons.

The Effective Lagrangian~\eqref{eq:left} is defined in the broken electroweak phase. However, NP scenarios can only be consistent with the direct search limits from the LHC if the new charged particles arise above the electroweak symmetry breaking scale. Therefore, to reinterpret our results for these scenarios, one should perform the renormalization group evolution from the low-energy scale $\mu_b$ up to $\mu_\mathrm{EW}\simeq m_W$~\cite{Jenkins:2017dyc}, and then match Eq.~\eqref{eq:left} to the so-called SMEFT (SM Effective Field Theory)~\cite{Grzadkowski:2010es,Buchmuller:1985jz}. The concrete ultraviolet scenario can then be matched to the SMEFT after accounting for the running effects above the electroweak scale $\mu_\mathrm{EW}$~\cite{Jenkins:2013zja}. Even though we present our results only in terms of the low-energy effective theory defined in Eq.~\eqref{eq:left}, we provide the needed inputs to recast our results to the most general NP scenario in Appendix~\ref{app:smeft}.

\section{$P\to P^\prime\ell\bar{\nu}$}
\label{sec:semileptonic}

\begin{figure}[t]
\centering
\begin{tikzpicture}
	\node at (0,0)[circle,fill,inner sep=1.5pt]{};
	\node at (0.25,-0.3) {$P$};
	\draw (0.5,2) -- (-3.5,2) -- (-4.5,-2) -- (-0.5,-2) -- (0.5,2);
	\draw[-stealth,decorate,decoration={snake,amplitude=3pt,pre length=2pt,post length=2pt}] (0,0) -- (-2,0);
 	\node at (-1,-0.5) {$\overrightarrow{q}$};
 	\draw[-stealth] (-2,0) -- (-1,1.5);
 	\draw[-stealth] (-2,0) -- (-3,-1.5);
 	\node at (2,0.3) {$\overrightarrow{u_z}$};
 	\node at (-1.75,1) {$\ell$};
 	\node at (-2.25,-1) {$\bar{\nu}$};
 	\draw[-stealth] (-2,0) ++ (.5,0) arc (0:64:.5 and .35);
 	\node at (-1.25,0.35) {$\theta_\ell$};
 	\draw[-stealth] (0,0) -- (2,0);
 	\node at (1,0.3) {$P^\prime$};
\end{tikzpicture}
\caption{\small \sl Angular convention for the process $P\to P^\prime \ell\nu$, where $P^{(\prime)}$ are pseudoscalar mesons. The angle  $\theta_\ell$ is defined in the rest frame of the meson $P$.}
\label{fig:semileptonic}
\end{figure}
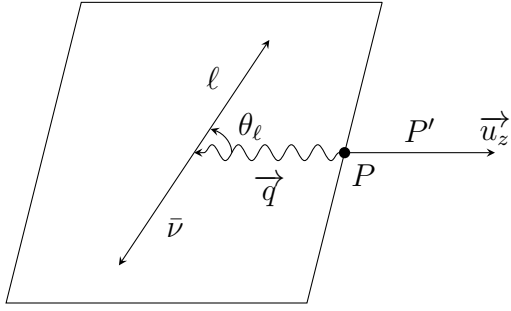

We first focus on $P\to P^\prime\ell \bar{\nu}$, where $P^{(')}$ denote the pseudoscalar mesons, for which one can build several observables that can be used to test the SM since the hadronic uncertainties in these modes are controlled by LQCD~\cite{Aoki:2019cca}. The differential $P\to P^\prime\ell \bar{\nu}$ decay distribution can be written in general as
\begin{align}
\label{eq:dB-general}
\dfrac{\mathrm{d}\mathcal{B}^\pm(q^2)}{\mathrm{d}q^2\, \mathrm{d}\cos\theta_\ell} = a^\pm(q^2) + b^\pm(q^2) \, \cos \theta_\ell + c^\pm(q^2)\, \cos^2 \theta_\ell\,,
\end{align}

\noindent where $q^2 = (p_\ell+p_\nu)^2$ with $m_\ell^2 < q^2\leq (m_P-m_{P^\prime})^2$, and $\theta_\ell$ is the angle between $\ell$ and the $P^\prime$ meson line-of-flight in the rest frame of the lepton pair, cf.~Fig.~\ref{fig:semileptonic}. The $\pm$ superscript stands for the polarization of the charged lepton, $\lambda_\ell$, and $a^\pm(q^2), b^\pm(q^2), c^\pm(q^2)$ are the $q^2$-dependent coefficients that are in principle sensitive to NP contributions.

The simplest observable, sensitive to the effective NP couplings, is the differential branching fraction,
\begin{align}
\begin{split}
\label{eq:BR-semileptonic}
\dfrac{\mathrm{d}\mathcal{B}(q^2)}{\mathrm{d}q^2} &= \int_{-1}^{1}\mathrm{d}\cos\theta_\ell \left[\dfrac{\mathrm{d}\mathcal{B}^+(q^2)}{\mathrm{d}q^2\, \mathrm{d}\cos\theta_\ell}+\dfrac{\mathrm{d}\mathcal{B}^-(q^2)}{\mathrm{d}q^2\, \mathrm{d}\cos\theta_\ell}\right]=2 \left[a(q^2) + \dfrac{c(q^2)}{3}\right]\,,
\end{split}
\end{align}
where $a(q^2) =a^+(q^2)+a^-(q^2)$, and $c(q^2)  =c^+(q^2) + c^-(q^2)$. This observable has already been copiously studied experimentally in the decays of $K$-, $D$- and $B$-mesons~\cite{Zyla:2020zbs}. The parameterization in Eq.~\eqref{eq:dB-general} suggests that there is more information that can be in principle extracted from these decays. To this purpose, one should further exploit the angular variables, as well as decays to the specifically polarized outgoing lepton. In the following, we show that four independent observables  can be defined and we provide their most general expressions.

\subsection{Form factors and helicity decomposition} The usual parameterization of the $P\to P^\prime \ell \bar{\nu}$ hadronic matrix elements reads 
\begin{align}
\label{eq:ff-vector}
 \langle P^\prime(k) |\bar{u}\gamma_\mu d | P(p) \rangle &= \bigg{[}(p+k)_\mu-\dfrac{M^2-m^2}{q^2}q_\mu 
 \bigg{]} f_+(q^2) + \dfrac{M^2-m^2}{q^2}q_\mu \,f_0(q^2)  \,, \\[0.4em]
 \label{eq:ff-tensor}
 \langle P^\prime(k) |\bar{u}\sigma_{\mu\nu} d | P(p) \rangle &= - i (p_\mu k_\nu-p_\nu k_\mu)\dfrac{2f_T(q^2,\mu)}{M+m} \,,
 \end{align}

\noindent where $f_{+,0,T}(q^2)$ are the hadronic form factors evaluated at $q^2=(p-k)^2$, while $M(m)$ denote the $P(P^\prime)$ meson masses. The relevant quark transition is denoted by $d\to u \ell\bar{\nu}$, where flavor indices are omitted for simplicity. The scalar matrix element can be obtained from Eq.~\eqref{eq:ff-vector} by using the Ward identity, which amounts to~\footnote{In the denominator of the right-hand-side of Eq.~\eqref{eq:SD} $m_d-m_u$ should be understood as the quark mass difference between the heavier and the lighter quarks. For instance for the $c\to d$ transition, $m_c-m_d$ should be in the denominator. For reference, we use the following quark mass values: $m_s^{\overline{\mathrm{MS}}}(2~\mathrm{GeV})=99.6(4.3)$~MeV, $m_c^{\overline{\mathrm{MS}}}(2~\mathrm{GeV})=1.176(39)$~GeV~\cite{Carrasco:2014cwa}, and $m_b^{\overline{\mathrm{MS}}}(m_b)=4.18(4)$~GeV~\cite{Zyla:2020zbs}.}
\begin{align}
\label{eq:SD}
 \langle P^\prime(k) |\bar{u}d | P(p) \rangle &= \dfrac{M^2-m^2}{m_d-m_u}\, f_0(q^2) \,.
 \end{align}

\noindent  With these definitions one can compute the coefficients $a^{\pm}(q^2)$, $b^{\pm}(q^2)$ and $c^{\pm}(q^2)$, defined in Eq.~\eqref{eq:dB-general}, as functions of the effective NP couplings, $g_\alpha^{ij\ell}$, introduced in Eq.~\eqref{eq:left}. To this purpose, it is convenient to perform a helicity decomposition of the decay amplitude by using the relation,
\begin{align}
\label{eq:pol-completeness}
\sum_{n, n^\prime}\varepsilon^{\ast\mu}_V (n)\varepsilon_V^\nu(n^\prime)g_{nn^\prime}=g^{\mu\nu}\,,
\end{align}
where $\varepsilon_V$ is the polarization vector of the virtual vector boson, as specified in Appendix~\ref{app:conventions}, with $n,n^\prime \in \lbrace {t,0 ,\pm}\rbrace$ and $g_{nn^\prime}=\mathrm{diag}(1,-1,-1,-1)$. The decay amplitude can then be decomposed in terms of the helicity amplitudes: 
\begin{align}
h_n(q^2) &= \varepsilon_V^{\mu\ast} (n)\,\left[ (1+g_V) \langle P^\prime | \bar{u}\gamma_\mu d| P \rangle + g_S\frac{q_\mu}{m_\ell}  \langle P^\prime | \bar{u}  d| P \rangle \right]
\,, \\[0.55em]
h_{nm}(q^2)& = \varepsilon_V^{\mu\ast} (n)\,\varepsilon_V^{\nu\ast} (m)\,g_T\,\langle P^\prime | \bar{u}\, i\sigma_{\mu\nu} d| P \rangle\,,
\end{align}

\noindent which are explicitly given by
\begin{align}
h_0(q^2) &= \left(1+g_V\right)\dfrac{\sqrt{\lambda(q^2,m^2,M^2)}}{\sqrt{q^2}}f_+(q^2)\,,\\[0.55em]
h_t(q^2) &= \left[1+g_V+g_S \dfrac{q^2}{m_\ell(m_d-m_u)}\right]\dfrac{M^2-m^2}{\sqrt{q^2}}f_0(q^2)\,,\\[0.55em]
h_{0t}(q^2) &= - h_{t0}(q^2) = - \,g_T\dfrac{\sqrt{\lambda(q^2,m^2,M^2)}}{m+M}f_T(q^2)\,,
\end{align}
where $\lambda(a^2,b^2,c^2)=[a^2-(b-c)^2][a^2-(b+c)^2]$. Other helicity amplitudes actually vanish. In order to express the physical observables defined in Eq.~\eqref{eq:dB-general} in a compact form, we define the following combination of helicity amplitudes
\begin{align}
h_0^{(+)}(q^2) &= h_0(q^2) - \frac{4\sqrt{q^2}}{m_\ell}h_{0t}(q^2)\,,\\[0.5em]
h_0^{(-)}(q^2) &= h_0(q^2) - \frac{4m_\ell}{\sqrt{q^2}}h_{0t}(q^2)\,,
\end{align}
which allows us to write
\begin{align}
\label{eq:a-coeff}
a^+(q^2) &= \mathcal{B}_0(q^2)\, m_\ell^2\, \big{|}h_t(q^2)\big{|}^2\,,\qquad\qquad\qquad\;\;\;\,
a^-(q^2) = \mathcal{B}_0(q^2)\, q^2 \, \big{|}h_0^{(-)}(q^2)\big{|}^2\,,\\[0.55em]
\label{eq:b-coeff}
b^+(q^2) &= \mathcal{B}_0(q^2)\, 2m_\ell^2\, \mathrm{Re}\big{[}h_0^{(+)}(q^2) h_t(q^2)^\ast\big{]}\,,\qquad b^-(q^2) = 0\,,\\[0.55em]
\label{eq:c-coeff}
c^+(q^2) &= \mathcal{B}_0(q^2)\, m_\ell^2\, \big{|}h_0^{(+)}(q^2)\big{|}^2\,,\;\qquad\qquad\qquad\, c^-(q^2) = -\mathcal{B}_0(q^2)\, q^2\, \big{|}h_0^{(-)}(q^2)\big{|}^2\,,
\end{align}

\noindent with 

\begin{equation}
\mathcal{B}_0 (q^2)= \tau_ P \, G_F^2 |V_{ij}|^2 \dfrac{\sqrt{\lambda(q^2,m^2,M^2)}}{256 \pi^3 M^3}\left(1-\dfrac{m_\ell^2}{q^2}\right)^2\,,
\end{equation}

\noindent where $\tau_P$ denotes the $P$-meson lifetime. From Eqs.~\eqref{eq:a-coeff} and \eqref{eq:c-coeff} we see that the following relations hold true,
\begin{equation}
\label{eq:relations-hel}
 b^-(q^2)=0\qquad\qquad\mathrm{and}\qquad\qquad a^-(q^2)=-c^-(q^2)\,.
\end{equation}
These equalities are respected not only in the SM, but also when the NP couplings are considered. Alternative way to derive the above expression is to make a partial-wave decomposition of the matrix elements, combined with selection rules for a left-handed neutrino. In other words, there are only four independent observables that can be constructed at the differential level, instead of six as one would naively infer from Eq.~\eqref{eq:BR-semileptonic}. These two relations could be a useful consistency check in experimental analyses in which the angular distribution to both polarization states of the charged-lepton are reconstructed. For decays to $\tau$ this is possible as the $\tau$-polarization can be reconstructed through its decay to one or three pions, for example. That methodology, however, cannot be applied to the decays to light leptons ($\mu$'s or $e$'s). 

\subsection{Physical observables} From the above discussion, we conclude that only four observables are linearly independent. We now list the set of observables which we will use in our subsequent phenomenological discussion. 

\begin{itemize}
 \item[\textbf{i)}] \emph{Branching fraction}: The first observable is the total branching fraction defined in Eq.~\eqref{eq:BR-semileptonic}, which is the most commonly considered in experimental searches, and which is given by 
 \begin{equation}
\mathcal{B}_\mathrm{tot} = \int_{m_\ell^2}^{(M-m)^2} \left( {d\mathcal{B}(q^2)\over d q^2}\right)\, dq^2\,,
\end{equation}
with $d\mathcal{B}(q^2)/d q^2$ already given in Eq.~\eqref{eq:BR-semileptonic}.
 
 \vspace*{0.4em}
 
 \item[\textbf{ii)}] \emph{Forward-backward asymmetry}: Another quantity that can be studied experimentally is the forward-backward asymmetry,
\begin{align}
\begin{split}
\label{eq:Afb}
\dfrac{\mathrm{d}A_{\mathrm{fb}}(q^2)}{\mathrm{d}q^2} &= \dfrac{1}{\mathcal{B}_\mathrm{tot}}\left[\int_{0}^{1}\mathrm{d}\cos\theta_\ell \dfrac{\mathrm{d}\mathcal{B}}{\mathrm{d}q^2\,\mathrm{d}\cos\theta_\ell}-\int_{-1}^{0}\mathrm{d}\cos\theta_\ell \dfrac{\mathrm{d}\mathcal{B}}{\mathrm{d}q^2\,\mathrm{d}\cos\theta_\ell}\right]=\dfrac{b(q^2)}{\mathcal{B}_\mathrm{tot}}\,,
\end{split}
\end{align}
where $\mathcal{B}=\mathcal{B}^+ + \mathcal{B}^-$ and $b(q^2)= b^+(q^2)+b^-(q^2)$, as defined above. This observable is normalized to the total branching fraction, $\mathcal{B}_\mathrm{tot}$. The above expression refers to the $q^2$-dependent quantity and its integrated characteristic is obtained after integration over the full $q^2$ range. 

 \vspace*{0.4em}

 \item[\textbf{iii)}] \emph{Lepton-polarization asymmetry}: A study of the decay to the charged lepton with a specific polarization state allows one to measure the lepton-polarization asymmetry defined as,
\begin{align}
\begin{split}
\dfrac{\mathrm{d}A_{\lambda}(q^2)}{\mathrm{d}q^2} &= \dfrac{1}{\mathcal{B}_\mathrm{tot}}\left[  \dfrac{\mathrm{d}\mathcal{B}^+}{\mathrm{d}q^2}- \dfrac{\mathrm{d}\mathcal{B}^-}{\mathrm{d}q^2}\right]\,,
\end{split}
\end{align}
 
  \noindent which depends on a complementary combination of helicity amplitudes, namely,
  
\begin{align}
\begin{split}
\label{eq:Apol}
\dfrac{\mathrm{d}A_{\lambda}(q^2)}{\mathrm{d}q^2} &=\dfrac{2}{\mathcal{B}_\mathrm{tot}} \left[a^+(q^2)-a^-(q^2) + \dfrac{1}{3}\Big{(}c^+(q^2)-c^-(q^2)\Big{)}\right]\,.
\end{split}
\end{align}

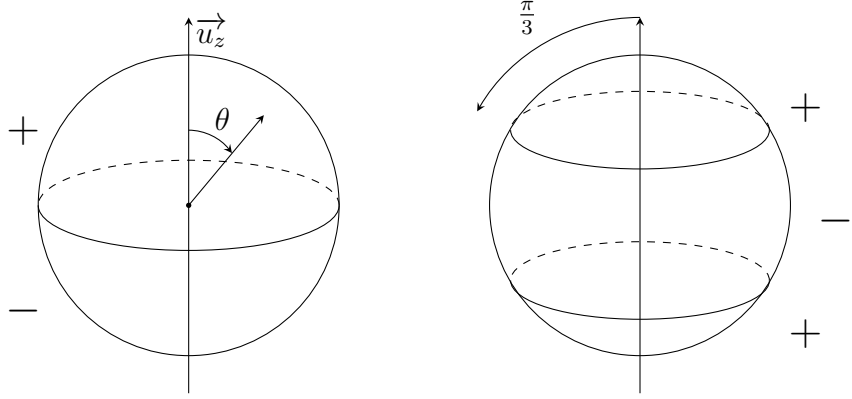
\begin{figure}[!t]
\centering
\begin{tikzpicture}
  \draw (0,0) circle (2);
  \draw (-2,0) arc (180:360:2 and 0.6);
  \draw[dashed] (2,0) arc (0:180:2 and 0.6);
  \fill[fill=black] (0,0) circle (1pt);
  \draw[-stealth] (0,-2.5)--(0,2.5);
  \draw[-stealth] (0,0)--(1,1.2);
  \draw[-stealth] (0,1) arc (90:44:0.8 and 1);
  \node at (0.45, 1.15) {$\theta$};
  \node at (0.3, 2.3) {$\overrightarrow{u_z}$};
  \node at (-2.2,1) {\Large $+$};
  \node at (-2.2,-1.4) {\Large $-$};
  \draw (6,0) circle (2);
  \draw (6-0.86*2,1) arc (180:360:0.86*2 and 0.86*0.6);
  \draw[dashed] (6+0.86*2,1) arc (0:180:0.86*2 and 0.86*0.6);
  \draw (6-0.86*2,-1) arc (180:360:0.86*2 and 0.86*0.6);
  \draw[dashed] (6+0.86*2,-1) arc (0:180:0.86*2 and 0.86*0.6);    
  \draw[-stealth] (6,-2.5)--(6,2.5);
  \draw[-stealth] (6,2.5) arc (90:150:2.5);
  \node at (4.5,2.5) {$\frac{\pi}{3}$};
  \node at (8.6,-0.2) {\Large $-$};
  \node at (8.2,-1.7) {\Large $+$};
  \node at (8.2,1.3) {\Large $+$};
\end{tikzpicture}
\caption{\small \sl Description to count the events for the angular asymmetry $A_\mathrm{fb}$ (left panel) and $A_{\pi/3}$ (right panel) as a function of the angle $\theta_\ell \in (0,\pi)$ defined in Fig.~\ref{fig:semileptonic}. Both observables are normalized to the total number of events.}
\label{fig:angular_observables}
\end{figure}

  \vspace*{0.4em}
 
  \item[\textbf{iv)}] \emph{Convexity}: The last independent observable that we consider is defined as follows,
\begin{align}
\begin{split}
\label{eq:Api3}
\dfrac{\mathrm{d}A_{\mathrm{\pi/3}}(q^2)}{\mathrm{d}q^2} = \dfrac{1}{\mathcal{B}_\mathrm{tot}} \bigg{[}\int_{1/2}^{1}\mathrm{d}\cos\theta_\ell \dfrac{\mathrm{d}\mathcal{B}}{\mathrm{d}q^2\mathrm{d}\cos\theta_\ell} &-\int_{-1/2}^{1/2}\mathrm{d}\cos\theta_\ell \dfrac{\mathrm{d}\mathcal{B}}{\mathrm{d}q^2\,\mathrm{d}\cos\theta_\ell}\\
&+\int_{-1}^{-1/2}\mathrm{d}\cos\theta_\ell \dfrac{\mathrm{d}\mathcal{B}}{\mathrm{d}q^2\,\mathrm{d}\cos\theta_\ell}\bigg{]}\,,
\end{split}
\end{align}
\noindent and allows us to single out the ``convexity'' coefficient $c(q^2)=c^+(q^2) + c^-(q^2)$\, i.e.,
\begin{align}
\dfrac{\mathrm{d}A_{\mathrm{\pi/3}}(q^2)}{\mathrm{d}q^2} &=\dfrac{c(q^2)}{2\,\mathcal{B}_\mathrm{tot}}\,.
\end{align}

\noindent While $A_\mathrm{fb}$ is defined as the symmetry between events collected in the regions $\theta\in (0, \pi/2)$ and $(\pi/2,\pi)$, the observable $A_{\mathrm{\pi/3}}$ measures the difference between events for which $\theta\in (\pi/3,2\pi/3)$ and those in the complementary angular region, as illustrated in Fig.~\ref{fig:angular_observables}.

\end{itemize}

\noindent In principle, one could define different set of observables but, as demonstrated in Eqs.~\eqref{eq:a-coeff}--\eqref{eq:c-coeff}, these observables would necessarily be a linear combination of the ones defined above. In other words, they do not provide us with any additional information on physics beyond the SM.

\section{$P\to \ell\bar{\nu}$ and $\ell\to P \nu$}
\label{sec:leptonic}

As far as the control of the underlying hadronic uncertainties is concerned, the leptonic decays of pseudoscalar mesons are among the cleanest probes of NP. The relevant hadronic matrix elements for these decays in the SM are defined as
\begin{align}
\label{eq:fp}
\langle 0 |\bar{u}\gamma^\mu \gamma_5 d| P(p) \rangle &= i f_P \,p^\mu\,,\end{align}
where $f_P$ is the $P$-meson decay constant. From Eq.~\eqref{eq:fp}, after applying the axial Ward identity, the matrix element of the pseudoscalar density reads
\begin{align}
 \langle 0 |\bar{u} \gamma_5 d| P(p) \rangle &= -i  \dfrac{f_P\,M^2}{m_u+m_d}\,,
\end{align}
which is also needed to describe the NP contributions. In other words the only hadronic quantity needed to describe the leptonic decay mode in the SM and its generic NP extension is the decay constant $f_P$. It is now straightforward to compute the branching fraction by using the effective Lagrangian~\eqref{eq:left}. We have,
\begin{align}
\label{eq:BRlep}
\mathcal{B}(P\to \ell \bar{\nu}) = \tau_P \dfrac{G_F^2 |V_{ij}|^2 f_P^2 M m_\ell^2}{8\pi} \left(1-\dfrac{m_\ell^2}{M^2}\right)^2 \left|1 - g_A + g_P \dfrac{M^2}{m_\ell (m_u+m_d)}\right|^2\,,
\end{align}

\noindent where $M$ and $\tau_P$ denote the  mass and the lifetime of $P$. We remind the reader that the effective coefficients $g_A$ and $g_P$ are related to the effective Lagrangian in Eq.~\eqref{eq:left} via the relations $g_A=g_{V_R}-g_{V_L}$ and $g_P = g_{S_R}-g_{S_L}$. For the $\tau$-lepton and light-quark transitions, it is the inverse process $\tau\to P \nu$ that is kinemetically available, $P=\pi^-, K^-$. These processes can also be computed in terms of $f_P$ and the effective NP couplings $g_{A,P}$,
\begin{align}
\label{eq:BRtaulep}
\mathcal{B}(\tau \to P {\nu}) = \tau_\tau \dfrac{G_F^2 |V_{ij}|^2 f_P^2 m_\tau^3}{16\pi} \left(1-\dfrac{M^2}{m_\tau^2}\right)^2 \left|1 - g_A - g_P \dfrac{M^2}{m_\ell (m_u+m_d)}\right|^2\,,
\end{align}
where $M$ denotes once again the $P$-meson mass.

\section{SM phenomenology}
\label{sec:numerics}

\subsection{Observables}

In order to reduce the theoretical uncertainties, we opt for building observables that are independent on the CKM matrix elements. These observables can be either a ratio of decays with distinct leptons in the final state, or a ratio of semileptonic and leptonic decays based on the same quark transition, as we describe in what follows.

\begin{itemize}
 \item \textbf{LFU ratios}: LFU ratios are powerful tests of validity of the SM, since both theoretical and experimental uncertainties cancel out in these ratios to a large extent. We define,
\begin{align}
\label{eq:ratio-lep}
R_{P}^{(\ell/\ell^\prime)} \equiv \dfrac{\mathcal{B}(P\to  \ell \bar{\nu})}{\mathcal{B}(P\to  \ell^\prime \bar{\nu})}\,,\qquad\qquad R_{PP^\prime}^{(\ell/\ell^\prime)} \equiv \dfrac{\mathcal{B}(P\to P^\prime \ell \bar{\nu})}{\mathcal{B}(P\to P^\prime \ell^\prime \bar{\nu})}\,,
\end{align}
where $P^{(\prime)}$ denotes a pseudoscalar meson and $\ell^{(\prime)}$ a charged lepton. Experimental results considered in our analysis are collected in Table~\ref{tab:observables}, along with the SM predictions that will be discussed in Section~\ref{ssec:hadronic-numerics}. SM predictions for leptonic decays have no uncertainty at leading order in QED, since the decay constant $f_P$ cancels out completely in Eq.~\eqref{eq:ratio-lep}. Moreover, the uncertainty of semileptonic ratios are rather small, since the normalization of $P\to P^\prime$ form factors cancels out in Eq.~\eqref{eq:ratio-lep}, while the remaining uncertainty from the form factor shapes is controlled by the LQCD results, as will be discussed in Sec.~\ref{ssec:hadronic-numerics}.

 \item \textbf{Semileptonic/leptonic ratios:} Another way to eliminate the dependence on the CKM matrix elements is to define the ratios,
\begin{align}
\label{eq:ratio-lep-semilep}
{r}_{PP^\prime}^{(\ell)} = \dfrac{\mathcal{B}(P^{\prime\prime}\to \ell {\nu})}{\mathcal{\overline{B}}(P \to P^{\prime} \ell {\nu})}\,,
\end{align}

\noindent where $P^{\prime\prime}\to \ell\bar{\nu}$ and $P\to P^\prime \ell\bar{\nu}$ are decays based on the same quark transition.~\footnote{Similar observables have defined for the $b\to u\ell \nu$ transition in Ref.~\cite{Banelli:2018fnx}.} The label in $r_{PP^\prime}^{(\ell)}$ refers to the mesons appearing in the semileptonic process, while $P^{\prime\prime}$ is uniquely fixed by the given transition. For instance, $P^{\prime\prime}=K$ for the kaon observables $r_{K\pi}^{(\ell)}$, which are based on the transition $s\to u \ell\nu$, and $P^{\prime\prime}=B_c$ for $r_{BD}^{(\ell)}$ and $r_{B_{s}D_{s}}^{(\ell)}$, which proceed via $b\to c \ell \nu$. The branching fraction in the denominator is defined by combining the semileptonic decays of neutral and charged mesons, as follows,
\begin{align}
\label{eq:isospin-average}
\mathcal{\overline{B}}(P^+ \to P^{\prime\,0} \ell \nu) \equiv \dfrac{1}{2}\left[ \mathcal{B}(P^+ \to P^{\prime\,0} \ell^+ \nu)+C_{P^{\prime \, 0}}^2\,\frac{\tau_{P^+}}{\tau_{P^0}}\,\mathcal{B}(P^0 \to P^{\prime\,+} \ell^- \bar{\nu}) \right]\,,
\end{align}
where $\tau_{P^{+}}(\tau_P^0)$ is the lifetime of the meson $P$ with electric charge $+1(0)$, and $C_{P^{\prime\,0}}$ is the Clebsch-Gordan coefficient, which is $1/\sqrt{2}$ for $P^\prime = \pi^0$ and $1$ otherwise, see e.g.~Eq.~\eqref{eq:kl3-corrections} below.~\footnote{For the decay modes such as $B_s\to D_s\ell \bar{\nu}$, where only one combination of electric charges is possible, the denominator in Eq.~\eqref{eq:ratio-lep-semilep} should be replaced by the standard branching fraction.}  The advantage of this definition is to combine meson decays with different lifetimes since the following relation holds, modulo small isospin-breaking corrections,
\begin{align}
\dfrac{\mathcal{B}(P^+ \to P^{\prime\,0} \ell^+ {\nu})}{\mathcal{B}(P^0 \to P^{\prime\,+} \ell^- \bar{\nu})} =  C_{P^{\prime\,0}}^2\,\dfrac{\tau_{P^+}}{\tau_{P^0}}\,.
\end{align}

The available experimental results for ${r}_{PP^\prime}^{(\ell/\ell^\prime)}$ are collected in Table~\ref{tab:observables-bis}, along with our SM predictions that will be discussed in Sec.~\ref{ssec:hadronic-numerics}. The relative hadronic uncertainty of the SM predictions is larger in this case compared to the LFU ratios, also listed in Table~\ref{tab:observables}, since they do not cancel out in the ratio. Nonetheless, the current level of accuracy of LQCD determinations for the relevant decay constants and form factors allow us to perform this type of study as well. Notably, these observables are complementary to  the ones defined above because they too are sensitive to the LFU contributions from NP which would normally cancel out in Eq.~\eqref{eq:ratio-lep}.

\end{itemize}

\begin{table}[t]
\renewcommand{\arraystretch}{1.8}
\centering
\begin{tabular}{|c|cc|}
\hline 
$f_P$ & Value$~[\mathrm{MeV}]$ & Ref.\\ \hline\hline
$f_\pi$ & $130.2(8)$ &   \cite{Aoki:2019cca}\\ 
$f_K$ & $155.7(3)$ & \cite{Aoki:2019cca}\\
$f_D$ & $212.0(7)$ & \cite{Aoki:2019cca} \\ 
$f_{D_s}$ & $249.9(5)$ & \cite{Aoki:2019cca} \\ 
$f_{B}$ & $190.0(1.3)$ & \cite{Aoki:2019cca} \\ 
$f_{B_c}$ & $434(15)$ & \cite{Colquhoun:2015oha} \\ \hline
\end{tabular}
\caption{\small \sl Decay constants obtained by numerical simulations of QCD on the lattice.}
\label{tab:decay-constants} 
\end{table}

\subsection{Hadronic inputs and SM predictions}
\label{ssec:hadronic-numerics}

In our analyses we use the LQCD results for hadronic inputs~\cite{Aoki:2019cca}. The decay constants used in this work are collected in Table~\ref{tab:decay-constants}, whereas 
the $P\to P^\prime$
form-factor parameterizations and the needed numerical inputs are summarized in Table~\ref{tab:references} (Appendix~\ref{app:FFinputs}). In our numerical analysis we will sample the fit parameters for each transition with a multivariate Gaussian distribution and the covariance matrices provided in the LQCD papers listed below.

\begin{itemize}
 \item $K\to \pi$: We use the $q^2$-shape of the $K\to\pi$ form factors $f_{0}(q^2)$ and $f_{+}(q^2)$ as reported in Ref.~\cite{Carrasco:2016kpy} from simulations with $N_f= 2+1+1$ dynamical quark flavors. Recently, the shapes of these form factors have also been determined in an independent LQCD study~\cite{Kakazu:2019ltq}, but from simulations with $N_f=2+1$ dynamical quarks. The results are fully compatible with those presented in Ref.~\cite{Carrasco:2016kpy}. Concerning the form factor normalization, i.e. $f_+(0)=f_0(0)$, we use the FLAG average~\cite{Aoki:2019cca},
\begin{equation}
 f_+(0)=0.9706(27)\,, 
 \end{equation}
 which is dominated by the results reported by MILC/Fermilab~\cite{Bazavov:2013maa} and by ETMC~\cite{Carrasco:2016kpy}. 
 As for the tensor form factor, the only available results come from Ref.~\cite{Baum:2011rm} which we will use in the following.
 \item $D\to \pi$ and $D\to K$: The scalar and vector form factors for $D\to\pi$ and $D\to K$ semileptonic decays have been computed in Ref.~\cite{Lubicz:2017syv} for all of the physically relevant $q^2$ values. Similar results for the tensor form factor, for both of these channels, have been presented in Ref.~\cite{Lubicz:2018rfs}.
 \item $B_{(s)}\to D_{(s)}$: The scalar and vector $B\to D$ form factors have been computed in Refs.~\cite{Lattice:2015rga} and \cite{Na:2015kha}, which are combined in our analysis. For the tensor form factor, we use the results for $f_T(q^2)/f_+(q^2)$ evaluated near the zero recoil in Ref.~\cite{Atoui:2013zza} and drive the ratio to low $q^2$ values by a small slope that we extracted from Ref.~\cite{Bernlochner:2017jka}.
 
 \item $B\to\pi$ and $B_s\to K$: The $B\to\pi$ scalar and vector form factors have been computed near zero recoil in Ref.~\cite{Lattice:2015tia,Flynn:2015mha} and combined in Ref.~\cite{Aoki:2019cca}, whereas the tensor one has been computed in Ref.~\cite{Bailey:2015nbd}. Similarly, the $B_s\to K$ scalar and vector form factors have been recently computed in Ref.~\cite{Bazavov:2019aom}. There are no available results for the tensor form factor but since the two decays are similar, we will assume that the ratio $f_T(q^2)/f_+(q^2)$ is the same for both channels, $B\to\pi\ell\bar{\nu}$ and $B_s\to K\ell\bar{\nu}$. Notice that these channels are particularly problematic due to a very large phase-space, which implies rather large theoretical uncertainties when extrapolating the LQCD results for form factors, which are available at large $q^2$'s, all the way down to $q^2\to 0$. For that reason, these decay modes will be discussed separately in Sec.~\ref{ssec:BtoPi}.
 
 \end{itemize}

\noindent For kaon decays it is also necessary to account for the subleading corrections in order to match both the experimental precision and the accuracy to which the hadronic matrix elements are evaluated in LQCD. Those subleading corrections are summarized in the following multiplicative factor\cite{Antonelli:2010yf},
\begin{equation}
\label{eq:kl3-corrections}
\mathcal{B}_{K_{\ell3}} \to \mathcal{B}_{K_{\ell3}} \, C_\pi^2 \, S_\mathrm{EW} \, \left(1 + \delta_\mathrm{em}^{K\ell}+\delta_{SU(2)}^{K\pi} \right)^2\,,
\end{equation}

\noindent where $S_\mathrm{EW}=1.0232(3)$ is the short-distance electroweak correction~\cite{Marciano:1993sh,Sirlin:1977sv}, $C_\pi$ is the Clebsch-Gordan coefficient ($1$ for decays to $\pi^\pm$ and $1/\sqrt{2}$ for those to $\pi^0$), while $\delta_\mathrm{em}^{K\ell}$ and $\delta_{SU(2)}^{K\pi}$ respectively stand for the channel-dependent electromagnetic and isospin-breaking corrections the values of which are given in Table~\ref{tab:kl3-corrections}. Very recently the first lattice QCD results of $\delta_\mathrm{em}^{K\ell}$ have been presented in Ref.~[30], and the reported values fully agree with those given in Table~\ref{tab:kl3-corrections}. Radiative corrections to $K_{\ell 2}$ have been estimated by using chiral perturbation theory (ChPT) and LQCD, leading to the SM prediction~\cite{Cirigliano:2007ga,DiCarlo:2019thl} 
\begin{equation}
\left(\dfrac{ \mathcal{B}_{K_{e2}} }{\mathcal{B}_{K_{\mu 2}}}\right)_\mathrm{SM} = 2.477(1)\times 10^{-5}\,.
\end{equation}
The electromagnetic correction to the muonic mode alone can be written as~\cite{Marciano:2004uf,Cirigliano:2011ny,Rosner:2015wva} 
\begin{equation}
\mathcal{B}_{K_{\mu 2}} \to \mathcal{B}_{K_{\mu 2}} (1+\delta_\mathrm{em}^{K_{\mu 2}})\,,
\end{equation}
\noindent where we take $\delta_\mathrm{em}^{K_{\mu 2}}= 0.0024(10)$, as recently determined in LQCD~\cite{DiCarlo:2019thl}. While the lattice determination of $\delta_\mathrm{em}^{\pi_{\mu 2}}$ appeared to be consistent with the one obtained in ChPT, the $\delta_\mathrm{em}^{K_{\mu 2}}$ value turned out to be much smaller than $\delta_\mathrm{em}^{K_{\mu 2}}=0.0107(21)$ as found in ChPT and previously used in phenomenology, cf. Ref.~\cite{Zyla:2020zbs} and references therein. As for the ratio of $\mathcal{B}_{K_{\mu 2}}$ and $\mathcal{B}_{\tau_{K 2}} \equiv \mathcal{B} (\tau \to K \bar{\nu})$, the radiative corrections are included by~\cite{Pich:2013lsa} 
\begin{equation}
\dfrac{\mathcal{B}_{\tau_{K 2}}}{\mathcal{B}_{K_{\mu_{2}}}}\to \dfrac{\mathcal{B}_{\tau_{K 2}}}{\mathcal{B}_{K_{\mu_{2}}}}\,(1+\delta R_{\tau/K})\,,
\end{equation}
\noindent with $\delta R_{\tau/K}=0.90(22)\times 10^{-2}$~\cite{Decker:1994ea}. For the observables related to the decays of $D_{(s)}$- and $B_{(s,c)}$-mesons, we do not include the electromagnetic corrections, because the evaluation of these effects is not available from theory yet. In the future, however, and with improved experimental and hadronic uncertainties, it will become necessary to account for these effects as well. Note in particular that  such effects are the leading theoretical uncertainties of the LFU ratios of leptonic decays, since the decay constants fully cancel out.~\footnote{Effects from soft-photon emission in semileptonic $B$-meson decays have been recently considered in Ref.~\cite{Bordone:2016gaq,deBoer:2018ipi,Isidori:2020acz}, see also~Ref.~\cite{Becirevic:2009fy}.}

\begin{table}[t]
\renewcommand{\arraystretch}{1.9}
\centering
\begin{tabular}{|c|cc|}
\hline 
Channel & $\delta_\mathrm{em}^{K\ell}\,\times 10^{-2}$ &  $\delta_{SU(2)}^{K\pi}\,\times 10^{-2}$ \\ \hline\hline
$K^0 \to \pi^+ e\bar{\nu}$ & $0.49 (11)$ & \multirow{2}{*}{$0$}  \\ 
$K^0 \to \pi^+ \mu\bar{\nu}$ & $0.70(11)$ &  \\ \hline
$K^+ \to \pi^0 e\bar{\nu}$ & $0.05(13)$ & \multirow{2}{*}{$2.9(4)$} \\ 
$K^+ \to \pi^0 \mu\bar{\nu}$ & $0.01(13)$ & \\ \hline
\end{tabular}
\caption{\small \sl  Summary of long-distance electromagnetic \big{(}$\delta_\mathrm{em}^{K\ell}$\big{)} and isospin breaking \big{(}$\delta_{SU(2)}^{K\pi}$\big{)} corrections for $K_{\ell3}$ decays \cite{Antonelli:2010yf}, see Eq.~\eqref{eq:kl3-corrections}.}
\label{tab:kl3-corrections} 
\end{table}

\begin{table}[p]
\centering
\vspace{-4.2em}
\scalebox{0.92}{ 
\renewcommand{\arraystretch}{1.5}
\begin{tabular}{|c|c|c|cc|}
\hline 
Observable & Definition & Our SM prediction & Exp.~value & Ref.\\ \hline\hline 
\rule{0pt}{1.95em}
$ R_{K^-\pi^0}^{(\mu/e)}$ & $\dfrac{\mathcal{B}(K^-\to\pi^0\mu\bar{\nu})}{\mathcal{B}(K^-\to\pi^0e\bar{\nu})}$ & $0.663(2)$ & $0.662(3)$ & \cite{Zyla:2020zbs} \\ 
\rule{0pt}{1.95em}
$R_{K_L\pi^\pm }^{(\mu/e)}$ & $\dfrac{\mathcal{B}(K_L\to\pi^\pm\mu\bar{\nu} )}{\mathcal{B}(K_L\to\pi^\pm e\bar{\nu})}$ & $0.666(2)$ & $0.666(4)$ & \cite{Zyla:2020zbs} \\ 
\rule{0pt}{1.95em}
$R_{K}^{(e/\mu)}$ & $\dfrac{\mathcal{B}(K^-\to e \bar{\nu})}{\mathcal{B}(K^-\to \mu \bar{\nu})}$ & $2.477(1)\times 10^{-5}$ & $2.488(9)\times 10^{-5}$ &  \cite{Zyla:2020zbs}\\ 
\rule{0pt}{1.95em}
$R_{K}^{(\tau/\mu)}$ & $\dfrac{\mathcal{B}(\tau \to K^- \bar{\nu})}{\mathcal{B}(K^-\to \mu \bar{\nu})}$  & $0.01126(3)$ & $0.0107(4)$ & \cite{Zyla:2020zbs}\\[.8em] \hline
\rule{0pt}{1.95em}
$R_{D^-\pi^0}^{(\mu/e)}$ & $\dfrac{\mathcal{B}(D^-\to\pi^0\mu\bar{\nu})}{\mathcal{B}(D^-\to\pi^0 e\bar{\nu})}$ & $0.9864(12)$ & $0.943(45)$ & \cite{Besson:2009uv,Ablikim:2017lks} \\ 
\rule{0pt}{1.95em}
$R_{D^0\pi^-}^{(\mu/e)}$ & $\dfrac{\mathcal{B}(D^0\to\pi^-\bar \mu {\nu})}{\mathcal{B}(D^0\to\pi^- \bar e {\nu})}$ & $0.9862(12)$ & $0.915(43)$ & \cite{Ablikim:2018frk} \\  
\rule{0pt}{1.95em}
$R_{D}^{(\mu/e)}$ & $\dfrac{\mathcal{B}(D^-\to\mu\bar{\nu})}{\mathcal{B}(D^-\to e\bar{\nu})}$ &  $4.24\times 10^4$ & $>42.5$ & \cite{Zyla:2020zbs} \\ 
\rule{0pt}{1.95em}
$R_{D}^{(\tau/\mu)}$ & $\dfrac{\mathcal{B}(D^-\to\tau\bar{\nu})}{\mathcal{B}(D^-\to \mu\bar{\nu})}$ & $2.67$ & $3.21(64)(43)$ & \cite{Ablikim:2019rpl}\\[.9em] \hline
\rule{0pt}{1.95em}
$R_{D^-K^0}^{(\mu/e)}$ & $\dfrac{\mathcal{B}(D^-\to {K^0}\mu\bar{\nu})}{\mathcal{B}(D^-\to {K^0}e\bar{\nu})}$ & $0.9751(10)$  & $1.003(25)$ & \cite{Zyla:2020zbs}\\ 
\rule{0pt}{1.95em}
$R_{D^0K^-}^{(\mu/e)}$ & $\dfrac{\mathcal{B}(D^0\to K^-\mu^+{\nu})}{\mathcal{B}(D^0\to K^-e^+{\nu})}$ &  $0.9751(10)$ & $0.973(14)$ &  \cite{Zyla:2020zbs}\\  
\rule{0pt}{1.95em}
$R_{D_s}^{(\mu/e)}$ & $\dfrac{\mathcal{B}(D_s\to \mu \bar{\nu})}{\mathcal{B}(D_s\to e \bar{\nu})}$ & $4.25\times 10^4$ & $>65.4$ & \cite{Zyla:2020zbs,Ablikim:2021hex} \\ 
\rule{0pt}{1.95em} 
$R_{D_s}^{(\tau/\mu)}$ & $\dfrac{\mathcal{B}(D_s\to \tau \bar{\nu})}{\mathcal{B}(D_s\to \mu \bar{\nu})}$ & $9.74$ & $10.0(5)$ & \cite{Zyla:2020zbs,Ablikim:2021hex} \\[.9em] \hline
\rule{0pt}{1.95em}
$R_{B}^{(\mu/e)}$ & $\dfrac{\mathcal{B}(B\to \mu \bar{\nu})}{\mathcal{B}(B\to e \bar{\nu})}$ & $4.27 \times 10^4$ & $>0.66$ & \cite{Zyla:2020zbs,Sibidanov:2017vph}\\
\rule{0pt}{1.95em}
$R_{B}^{(\tau/\mu)}$ & $\dfrac{\mathcal{B}(B\to \tau \bar{\nu})}{\mathcal{B}(B\to \mu \bar{\nu})}$ & $2.23\times 10^2$ & $1.7(8)\times 10^2$ & \cite{Zyla:2020zbs,Sibidanov:2017vph}\\[.9em] \hline 
\rule{0pt}{1.95em}
$R_{BD}^{(\mu/e)}$ & $\dfrac{\mathcal{B}(B\to D \mu \bar{\nu})}{\mathcal{B}(B\to D e \bar{\nu})}$ & $0.9960(2)$ & $0.995(22)(39)$ & \cite{Glattauer:2015teq} \\
\rule{0pt}{1.95em}
$R_{B_sD_s}^{(\mu/e)}$ & $\dfrac{\mathcal{B}(B_s\to D_s \mu \bar{\nu})}{\mathcal{B}(B_s\to D_s e \bar{\nu})}$ & $0.9960(2)$  & -- & \\
\rule{0pt}{1.95em}
$R_{BD}^{(\tau/\mu)}$ & $\dfrac{\mathcal{B}(B\to D \tau \bar{\nu})}{\mathcal{B}(B\to D \mu \bar{\nu})}$ & $0.295(6)$ & $0.340(27)(13)$ &  \cite{Amhis:2019ckw} \\
\rule{0pt}{1.95em}
$R_{B_sD_s}^{(\tau/\mu)}$ & $\dfrac{\mathcal{B}(B_s\to D_s\tau\bar{\nu})}{\mathcal{B}(B_s\to D_s\mu\bar{\nu})}$ & $0.295(6)$ & -- &  \\[.9em] \hline 
\end{tabular}
}
\caption{\small \sl Experimental results for LFU ratios and SM predictions obtained by using the hadronic inputs described in Sec.~\ref{ssec:hadronic-numerics}. Ratios with semileptonic $B\to \pi(K)$ decays are discussed in Sec.~\ref{ssec:BtoPi}. When quoted, first uncertainty corresponds to the statistical and second to systematic. Upper limits are displayed at $90\%$ C.L. }
\label{tab:observables} 
\end{table}

With the ingredients described above, we are able to make the SM predictions that are listed in Table~\ref{tab:observables} and \ref{tab:observables-bis} for the two types of observables that we consider: (i) LFU tests, and (ii) ratios of semileptonic and leptonic decays, based on the same weak process. We find a reasonable agreement between our predictions and the experimental results, with a few exceptions which will be mentioned in the following.

\subsection{Discussion}
\label{ssec:discussion}

\paragraph{$K\to l \nu$, $K\to \pi l \nu$ and $|V_{us}|$:} In the kaon sector, we find a good agreement between the SM predictions and experiment for the LFU, as it can be seen in Table~\ref{tab:observables}. For the ratios of leptonic and semileptonic decays we find a reasonable agreement for the electron modes, while for the muonic modes we see a clear discrepancy.  More specifically, the SM prediction and the experimental values differ by $3.1\,\sigma$:
\begin{equation}
\label{eq:kaon-lep-semilep-sm-value}
\frac{ \mathcal{B}(K^-\to \mu  \nu)^\mathrm{SM} }{ \overline{ \mathcal{B}}(K^-\to \pi^0 \mu  \bar{\nu} )^\mathrm{SM} } = 18.55(16),
 \qquad\qquad  \dfrac{\mathcal{B}(K^-\to \mu  \nu)^\mathrm{exp}}{\overline{ \mathcal{B}}(K^-\to \pi^0 \mu  \bar{\nu})^\mathrm{exp}}=19.16(11)\,,
\end{equation}
where in the denominator we use the isospin average according to Eq.~\eqref{eq:isospin-average}. Also taken separately (without the isospin averaging), the measured values of the ratios are larger than the ones predicted in the SM:
\begin{align}
\label{eq:kaon-lep-semilep-sm-value2}
&\dfrac{\mathcal{B}(K^-\to \mu  \nu)^\mathrm{SM}}{ \mathcal{B}(K^-\to \pi^0 \mu  \bar{\nu})^\mathrm{SM}} = 18.26(17),\qquad\qquad  &\dfrac{\mathcal{B}(K^-\to \mu  \nu)^\mathrm{exp}}{ \mathcal{B}(K^-\to \pi^0 \mu  \bar{\nu})^\mathrm{exp}}&=18.9(2)\,,\nn\\ 
& \dfrac{\mathcal{B}(K^-\to \mu  \nu)^\mathrm{SM}}{ \mathcal{B}(K_L\to \pi^+ \mu  \bar{\nu})^\mathrm{SM}} = 2.28(2),\qquad\qquad  &\dfrac{\mathcal{B}(K^-\to \mu  \nu)^\mathrm{exp}}{ \mathcal{B}(K_L\to \pi^+ \mu  \bar{\nu})^\mathrm{exp}}&=2.352(11)\,.
\end{align}

Another way to see that problem has been already pointed out when extracting the value of $|V_{us}|$ from leptonic and semileptonic decay respectively~\cite{Moulson:2017ive}. We get: 
\begin{align}
\label{eq:vus-numerics}
 |V_{us}|_{K_{\mu 2}}  = 0.2264(6)\,,\qquad\qquad 
  |V_{us}|_{K_{\mu 3}} &= 0.2228(8)\,,
 \end{align}
with the latter value fully compatible with the one extracted from the electronic mode, $|V_{us}|_{K_{e 3}} = 0.2228(7)$. Clearly, the two values in Eq.~\eqref{eq:vus-numerics} differ by $3.5\,\sigma$. Understanding the origin of that discrepancy requires a proper assessment of the electromagnetic corrections entering the expressions for the $K_{\ell 3}$ decays by means of LQCD. 

As a side exercise, one can use the ratio of the accurately measured leptonic decays $K_{\mu 2}/\pi_{\mu 2}$, for which the electromagnetic corrections have been handled by LQCD~\cite{Giusti:2017dwk}, and combine it with the ratio of decay constants $f_K/f_\pi = 1.193(2)$~\cite{Aoki:2019cca}. As a result we get $|V_{us}|/|V_{ud}| = 0.2319(5)$. If we neglect $|V_{ub}|$ and impose the CKM unitarity we obtain~\footnote{Note that the value of $|V_{ub}|$ is irrelevant for this discussion since its central value is too small compared to the current precision in the determination of $|V_{us}|$ and $|V_{ud}|$.} 
\bea
|V_{us}|_{ K_{\mu 2}/\pi_{\mu 2} }^\mathrm{CKM}  = 0.2259(5)\,.
\eea
The same value is obtained if instead of invoking the CKM unitarity we multiply $|V_{us}/V_{ud}|_{K_{\mu 2}/\pi_{\mu 2}}$ by $|V_{ud}|_\beta$, extracted from the nuclear $\beta$-decay~\cite{Czarnecki:2019mwq}  (see also Ref.~\cite{Marciano:2005ec,Seng:2018yzq,Seng:2020wjq,Crivellin:2020lzu} and references therein). 
These values are clearly in good agreement with $|V_{us}|_{K_{\mu 2}}$, but not with $|V_{us}|_{K_{\mu 3}}$. Moreover, the discrepancy between $|V_{us}|_{K_{\mu 3}}$ and $|V_{us}|_{K_{\mu 2}}$ is larger if considering the semileptonic decays of charged kaons.

In short, an improved LQCD determination of the $K\to\pi$ form factors, and especially a good control over the electromagnetic corrections is needed in order to clarify this discrepancy. If this discrepancy persists then a viable NP explanation would necessitate introducing the LFU couplings in order to guarantee a consistency with $R_{K\pi}^{(\mu/e)}$, where the SM predictions and the experimental measurements agree very well, cf.~Table~\ref{tab:observables}. 

Before closing this discussion we should emphasize the fact that for the semileptonic decays we took the values for $\mathcal{B}(K^-\to \pi^0 l\bar \nu)^\mathrm{exp}$ from Ref.~\cite{Moulson:2017ive}. Had we used the simple averages of the measurements reported in the literature, and listed in PDG Review~\cite{Zyla:2020zbs}, the abovementioned discrepancy between $|V_{us}|_{K_{\mu 2}}$ and $|V_{us}|_{K_{\ell 3}}$ would increase to $5\,\sigma$. We believe that more discussion in assessing the correct values of the experimental branching fractions in the kaon decays is needed. For example, the value of $\mathcal{B}(K^-\to \pi^0\mu \bar \nu)^\mathrm{exp} =3.366(30)~\%$ as suggested in Ref.~\cite{Moulson:2017ive} is very close to the value reported in the PDG Review as ``Our Fit", but it is $2.7\, \sigma$ larger than the ordinary average which is heavily dominated by the result reported by the KLOE collaboration, namely $\mathcal{B}(K^-\to \pi^0\mu \bar \nu)^\mathrm{exp} =3.233(39)~\%$~\cite{Ambrosino:2007aa}. Similar situation is true for $\mathcal{B}(K^-\to \pi^0e \bar \nu)^\mathrm{exp}$.
\begin{table}[t]
\renewcommand{\arraystretch}{1.9}
\centering
\vspace{-3em}
\begin{tabular}{|c|c|c|cc|}
\hline 
Observable & Definition & Our SM prediction & Exp.~value & Ref.\\ \hline\hline 
\rule{0pt}{2.1em}
$ r_{K\pi}^{(e)}$ & $\dfrac{\mathcal{B}(K^-\to e \bar{\nu})}{\mathcal{\overline{B}}(K^- \to \pi^0 e \bar{\nu})}$ &$3.05(3)\times 10^{-4}$ & $3.17(2)\times 10^{-4}$ & \cite{Moulson:2017ive} \\ 
\rule{0pt}{2.1em}
$ r_{K\pi}^{(\mu)}$ & $\dfrac{\mathcal{B}(K^-\to \mu \bar{\nu})}{\mathcal{\overline{B}}(K^- \to \pi^0 \mu \bar{\nu})}$ &  $18.6(2)$ & $19.2(1)$ & \cite{Moulson:2017ive} \\[.8em] \hline 
\rule{0pt}{2.1em}
$ r_{D\pi}^{(e)}$ & $\dfrac{\mathcal{B}(D^-\to e \bar{\nu})}{\mathcal{\overline{B}}(D^- \to {\pi}^0 e \bar{\nu})}$ & $2.79(12)\times 10^{-6}$ & $< 2.4\times 10^{-3}$ & \cite{Zyla:2020zbs} \\ 
\rule{0pt}{2.1em}
$ r_{D\pi}^{(\mu)}$ & $\dfrac{\mathcal{B}(D^-\to \mu \bar{\nu})}{\mathcal{\overline{B}}(D^- \to \pi^0 \mu \bar{\nu})}$ & $0.120(5)$ &  $0.108(7)$ & \cite{Zyla:2020zbs} \\[.8em] \hline 
\rule{0pt}{2.1em}
$ r_{DK}^{(e)}$ & $\dfrac{\mathcal{B}(D_s \to e  \bar{\nu})}{\mathcal{\overline{B}}(D^- \to K^0 e {\bar \nu})}$ &$1.41(7)\times 10^{-6}$ & $<9\times 10^{-4}$ & \cite{Zyla:2020zbs} \\ 
\rule{0pt}{2.1em}
$ r_{DK}^{(\mu)}$ & $\dfrac{\mathcal{B}(D_s \to \mu \bar{\nu})}{\mathcal{\overline{B}}(D^- \to K^0 \mu {\bar \nu})}$ &$0.061(2)$ &  $0.063(2)$ & \cite{Zyla:2020zbs,Ablikim:2021hex} \\[.8em] \hline 
\rule{0pt}{2.1em}
$ r_{B D}^{(\mu)}$ & $\dfrac{\mathcal{B}(B_c^-\to \mu \bar{\nu})}{\overline{\mathcal{{B}}}(B^-\to D^0\mu{\bar \nu})}$ &$4.3(4)\times 10^{-3}$ & -- & \\
\rule{0pt}{2.1em}
$ r_{B D}^{(\tau)}$ & $\dfrac{\mathcal{B}(B_c^-\to \tau\bar{\nu})}{\overline{\mathcal{B}}(B^-\to D^0\tau\bar{\nu})}$ & $3.5(3)$ & -- & \\[1.em] \hline 
\end{tabular}
 \caption{\small \sl Experimental results for ratios of leptonic and semileptonic decays, and SM predictions obtained by using the hadronic inputs described in Sec.~\ref{ssec:hadronic-numerics}. Ratios with semileptonic $B\to \pi (K)$ decays are discussed in Sec.~\ref{ssec:BtoPi}}
\label{tab:observables-bis} 
\end{table}

\begin{figure}[t!]
\centering
\includegraphics[width=0.5\linewidth]{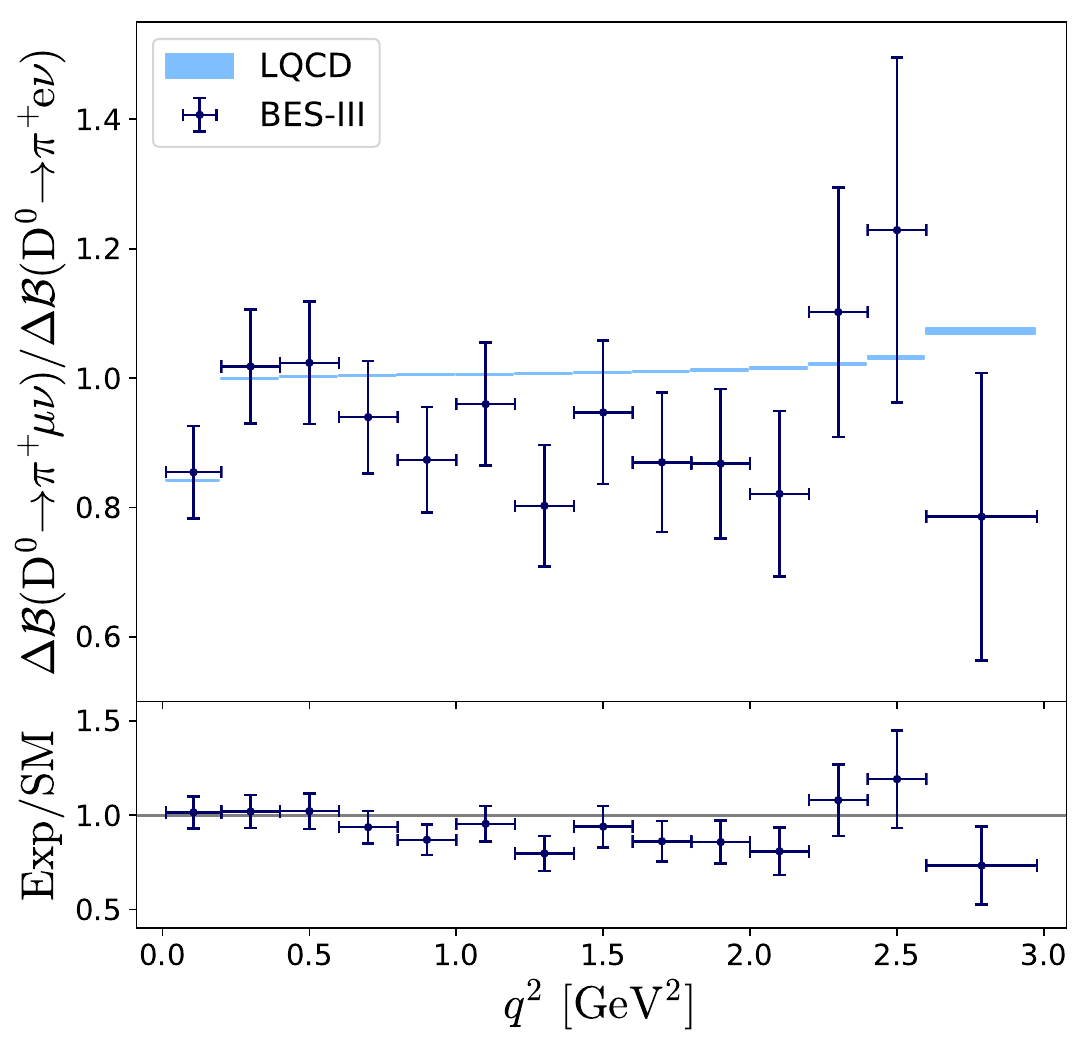}~\includegraphics[width=0.5\linewidth]{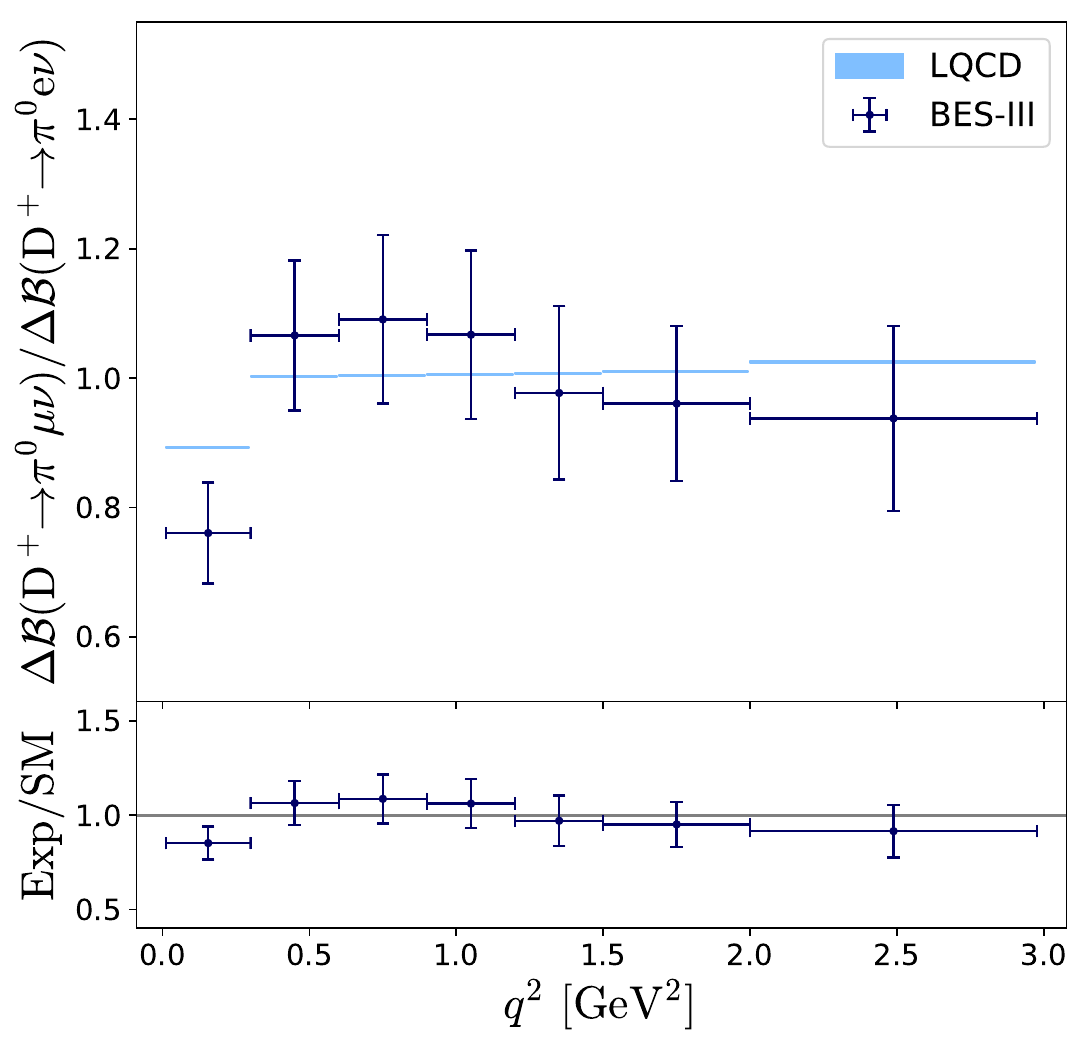}
\caption{\small \sl Comparison between the $\mu/e$ LFU ratios measured experimentally in different $q^2$ bins for $D^0\to \pi^+ l \nu$~\cite{Ablikim:2018frk,Ablikim:2015ixa} (left panel) and $D^+\to \pi^0 l \nu$~\cite{Ablikim:2017lks,Ablikim:2018frk} (right panel) with the SM predictions (shaded blue regions).}
\label{fig:dB-DPilnu}
\end{figure}

\paragraph{$D\to \pi l \nu$ and $|V_{cd}|$:} As it can be seen in Table~\ref{tab:observables}, we also find mild discrepancies between theory and experiment in the $D\to\pi\l\bar \nu$. These are mostly related to the recent BES-III results on $D^0\to\pi^+ l\bar{\nu}$ decays (with $l=e,\mu$)~\cite{Ablikim:2018frk,Ablikim:2015ixa}. To investigate this problem, we compare in Fig.~\ref{fig:dB-DPilnu} the ratio of the $D\to\pi \mu\bar{\nu}$ and $D\to\pi e\bar{\nu}$ differential distributions measured experimentally for both $D^+$ and $D^0$ decays~\cite{Ablikim:2018frk,Ablikim:2015ixa,Ablikim:2017lks} with the SM predictions based on the form factors taken from Ref.~\cite{Lubicz:2017syv}. While there is a good agreement between theory and experiment for $D^+\to\pi^0 l\bar{\nu}$ decays, we observe mild discrepancies in several $q^2$ bins of $D^0\to\pi^+ l\bar{\nu}$ (see also Ref.~\cite{Riggio:2017zwh}). Since these deviations only appear in one of the decay modes, it is likely that they arise from an underestimated theoretical or experimental uncertainty near the zero recoil. In other words, most NP scenarios would not be able to explain this discrepancy since they would contribute equally to the both decay modes. Note that these observables have recently been analyzed in a similar context in Refs.~\cite{Fleischer:2019wlx,Leng:2020fei}.

\begin{figure}[t!]
\centering
\includegraphics[width=0.5\linewidth]{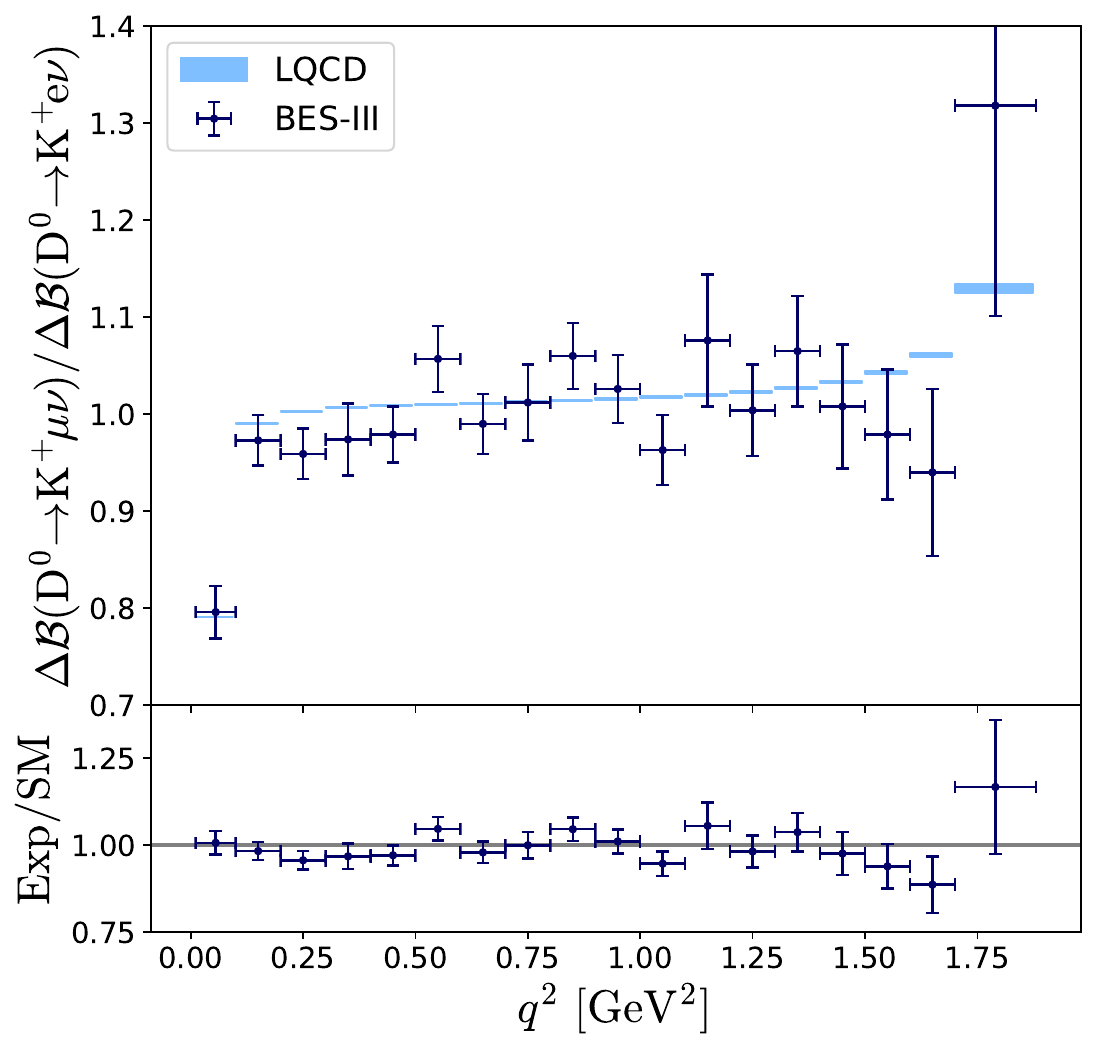}
\caption{\small \sl Comparison between the $\mu/e$ LFU ratios measured experimentally in different $q^2$ bins for $D^0\to K^+ \ell \nu$~\cite{Ablikim:2015ixa,Ablikim:2018evp} with the SM predictions (shaded blue regions). The isospin-related decay modes $D^+\to K^0 \ell \nu$ are not shown since the differential data for $D^+\to K^0 \mu \nu$ is not available~\cite{Ablikim:2016sqt}.}
\label{fig:dB-DKlnu}
\end{figure}

\paragraph{$D\to K l \nu$ and $|V_{cs}|$:} For the $D\to K$ transition we find a reasonable agreement between theory and experiment. This conclusion is true for both LFU tests, as it can be seen in Table~\ref{tab:observables} and Table~\ref{tab:observables-bis}. The plot analogous to those discussed in the $D\to \pi$ case is shown in Fig.~\ref{fig:dB-DKlnu}. We observe a good agreement between the SM predictions and the measured LFU ratios in most of the $q^2$-bins.~\footnote{See Ref.~\cite{Fajfer:2015ixa} for a recent study of the related decay mode $D_s\to\phi \ell \bar{\nu}$ with lattice QCD form factors~\cite{Donald:2013pea}.}

\paragraph{$B\to D\ell\nu$ and LFU violation:} Lastly, there are hints of LFU violation in the $b\to c\tau \bar{\nu}$ transition. These deviations appear not only in the ratio $R_{BD}^{(\tau/\mu)}$, that shows an $\approx 1.5\sigma$ excess with respect to the SM prediction (cf.~Table~\ref{tab:observables})~\cite{Lees:2012xj,Huschle:2015rga}, but also in the related decay modes, $B\to D^\ast \ell \bar{\nu}$~\cite{Lees:2012xj,Huschle:2015rga,Hirose:2016wfn,Aaij:2015yra} and $B_c\to J/\psi \ell \bar{\nu}$~\cite{Aaij:2017tyk}, which are $\approx 2.5\sigma$ and $\approx 2\sigma$ above the corresponding SM predictions respectively. This pattern of deviations has triggered an intense activity in the theory community which resulted in several viable scenarios beyond the SM capable of accommodating the so-called $B$-anomalies (see e.g.~Ref.~\cite{Angelescu:2018tyl,Buttazzo:2017ixm} and references therein). The SM predictions for the $B\to D^\ast$ transition are currently made by relying on the differential distributions measured experimentally for $B\to D^\ast (\to D\pi) l \bar{\nu}$ decays (with $l=e,\mu$)~\cite{Amhis:2019ckw}, as well as the heavy-quark effective theory combined with the QCD sum rules to evaluate the non-perturbative coefficients entering the heavy quark expansion of the form factors, and in particular to evaluate the pseudoscalar form factor~\cite{Bernlochner:2017jka}. Although the LQCD results at nonzero recoil are not yet available for this particular transition, there are ongoing lattice studies the results of which will be helpful in clarifying the situation, and hopefully in understanding the long-standing disagreement between the $|V_{cb}|$ values as inferred from the exclusive and inclusive semileptonic decays, respectively~\cite{Gambino:2020jvv}. For the $B_c\to J/\psi$ transition, the relevant form factors at nonzero recoil have been recently computed by means of LQCD simulations in Ref.~\cite{Harrison:2020gvo}, which allows us to predict the corresponding LFU ratio $R_{B_c J/\psi}^{(\tau/\mu)}$, with $\mathcal{O}(1\%)$ precision~\cite{Harrison:2020nrv}, see also Ref.~\cite{Isidori:2020eyd}.

\begin{figure}[t!]
\centering
\includegraphics[width=0.5\linewidth]{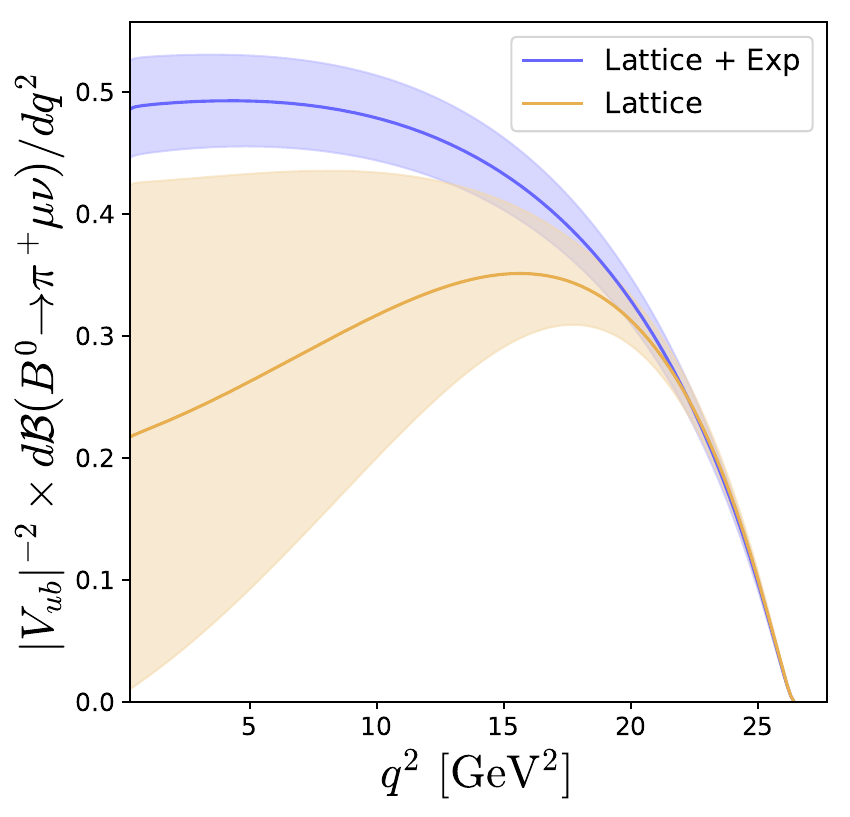}~\includegraphics[width=0.5\linewidth]{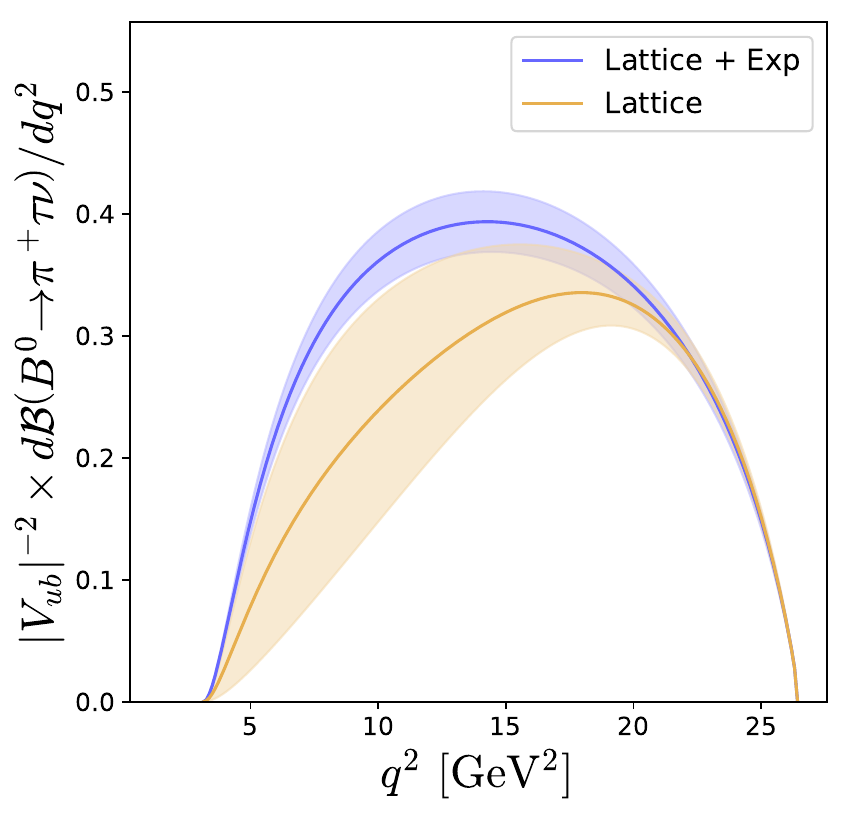}
\caption{\small \sl Differential branching fraction for $B\to \pi \mu \bar{\nu}$ (left panel) and $B\to\pi\tau\bar{\nu}$ (right panel) by using only LQCD form factors (orange)~\cite{Lattice:2015tia,Flynn:2015mha}, and a combined fit to LQCD and experimental data (blue)~\cite{Aoki:2019cca}. The shaded regions correspond to the $1\sigma$ predictions.}
\label{fig:dB-BPilnu}
\end{figure}

\subsection{$B\to \pi\ell\nu$ with LQCD form factors}
\label{ssec:BtoPi} 

The $B\to \pi\ell\nu$ and $B_s\to K\ell\nu$ decays deserve a separate discussion due to the large theoretical uncertainties involved in their SM predictions. For these processes, the form factors obtained in LQCD simulations at large $q^2$'s should be extrapolated to lower $q^2$'s in order to cover the entire physical region. This extrapolation introduces an additional source of uncertainty related to various parameterizations one might use to describe the $q^2$ dependencies of the form factors. In principle, this issue could be avoided by combining the lattice data with experimental data which are more accurate for low $q^2$'s, but that would be at odds with our goal to solely rely on LQCD to evaluate the hadronic matrix elements. Moreover for our purpose it is important to avoid using the  experimental data to constrain the form factors because such the results could already be heavily affected by the NP contributions which we would like to isolate.

\begin{figure}[t!]
\centering
\includegraphics[width=0.5\linewidth]{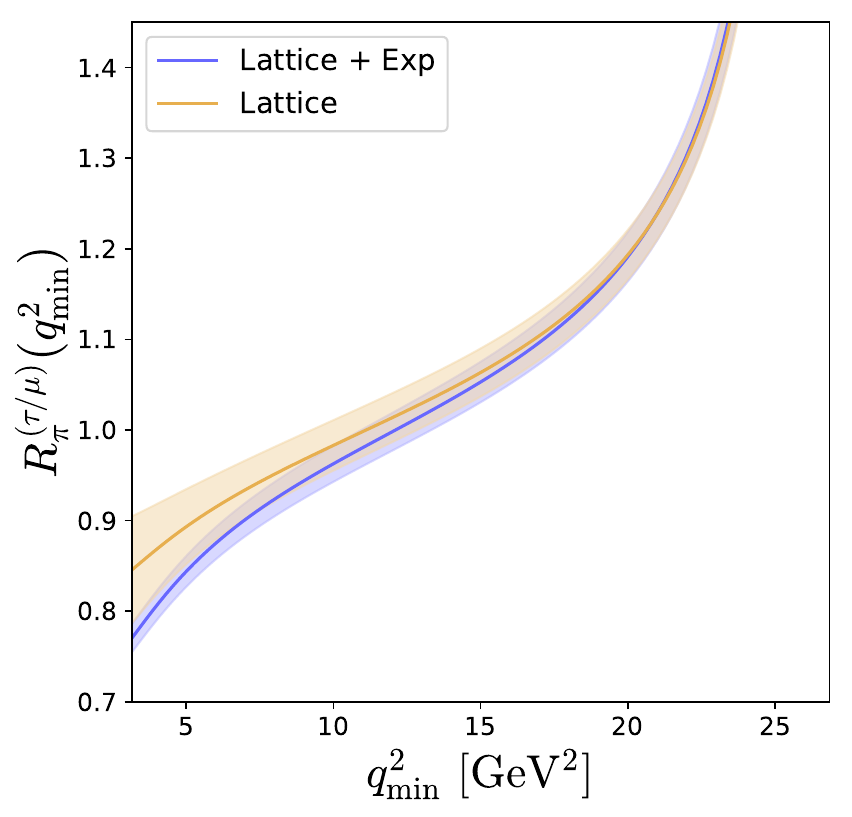}
\caption{\small \sl The ratio $\mathcal{R}_{B \pi}^{(\tau/\mu)}(q^2_\mathrm{min})$, defined in Eq.~\eqref{eq:Rcut}, is plotted as a function of the mininum value of the dilepton mass, $q^2_\mathrm{min}$, which is taken to be the same in the numerator and denominator.}
\label{fig:Rpi-cut}
\end{figure}

The uncertainty related to the form factor parameterization is noticeable for $B\to\pi\ell\nu$ decays, see e.g. Ref.~\cite{Bernlochner:2015mya}. In Fig.~\ref{fig:dB-BPilnu} we compute the  $B\to\pi\ell\nu$ differential decay rates by using two different theoretical inputs: (i) the scalar and vector form factors computed on the lattice at high-$q^2$ values and extrapolated to the rest of the physical region~\cite{Lattice:2015tia,Flynn:2015mha} (see Table~\ref{tab:references} in Appendix~\ref{app:FFinputs}); and (ii) $f_{0}(q^2)$ and $f_{+}(q^2)$ obtained by a combined fit of LQCD data with the experimental measurements of $\mathrm{d}\mathcal{B}(B\to \pi l \nu)/\mathrm{d}q^2$ (with $l=e,\mu$), which are more accurate at low $q^2$-values~\cite{Aoki:2019cca}.~\footnote{For reference, the numerical inputs needed to reproduce these form factors are given in Table 50 of Ref.~\cite{Aoki:2019cca}.} Note, in particular, that the second approach allows us to extract $|V_{ub}|=3.73(14)\times 10^{-3}$~\cite{Aoki:2019cca}, lower than the one extracted from the inclusive decays (see e.g.~Ref.~\cite{Gambino:2020jvv} for a recent review). Our predictions by using both sets of form factors are shown in Fig.~\ref{fig:dB-BPilnu}. Both approaches lead to the same results in the large $q^2$-region where LQCD data dominate, but they diverge for small $q^2$ values, due to the model dependent extrapolation of the LQCD form factors. The LFU ratios defined in Eq.~\eqref{eq:ratio-lep} are then~\footnote{Note that a similar problem is not present in the $\mu/e$ ratios, since the form factors cancel out to a large extent in these observables because $m_{e}\ll m_{\mu}\ll m_B$.}
\begin{equation}
 R_{B\pi}^{(\tau/\mu)}\Big{\vert}_\mathrm{LQCD} =0.78(10)\,, \qquad\qquad   R_{B\pi}^{(\tau/\mu)}\Big{\vert}_{\mathrm{LQCD}+\mathrm{exp}} =0.66(2)\,.
\end{equation}
Therefore, it is still not possible to use only LQCD data and have a robust SM prediction for $R_{B\pi}^{(\tau/\mu)}$. To avoid the artifact of the form factor extrapolations, we propose to use, instead of Eq.~\eqref{eq:ratio-lep}, the following observable,~\footnote{A similar proposal has been recently made for the $P\to V \ell \bar{\nu}$ transitions in Ref.~\cite{Isidori:2020eyd}, where $V$ denotes a vector meson. In this case, the uncertainties related to the pseudoscalar form factor can be substantially reduced by increasing the value of $q^2_\mathrm{min}$.}
\begin{equation}
\label{eq:Rcut}
\widehat{R}_{P P^\prime}^{(\ell/\ell^\prime)}(q^2_\mathrm{min}) \equiv  \dfrac{\displaystyle\int_{q_\mathrm{min}^2}^{(M-m)^2} \dfrac{\mathrm{d}\mathcal{B}}{\mathrm{d}q^2} (P\to P^\prime \ell \bar{\nu})\, \mathrm{d} q^2  }{\displaystyle\int_{q_\mathrm{min}^2}^{(M-m)^2} \dfrac{\mathrm{d}\mathcal{B}}{\mathrm{d}q^2} (P\to P^\prime \ell^\prime \bar{\nu}) \,\mathrm{d} q^2  }\,,
\end{equation}

\begin{table}[t!]
\renewcommand{\arraystretch}{2.0}
\centering
\begin{tabular}{|c|c|cc|}
\hline
Observable & Our SM prediction & Exp.~value & Ref. \\ \hline\hline
$ \widehat{R}_{B\pi}^{(\mu/e)} (16~\mathrm{GeV}^2)$ & $1.0007(1)$ & -- &  \\ 
$ \widehat{R}_{B_s K}^{(\mu/e)} (16~\mathrm{GeV}^2)$ & $1.0009(1)$ & -- & \\ 
$ \widehat{R}_{B\pi}^{(\tau/\mu)} (16~\mathrm{GeV}^2)$ & $1.08(3)$ & $<6.4$ & \cite{Lees:2012vv,Sibidanov:2013rkk,Hamer:2015jsa} \\ 
$ \widehat{R}_{B_s K}^{(\tau/\mu)} (16~\mathrm{GeV}^2)$ & $1.10(2)$ & -- & \\ \hline\hline
$ \widehat{r}_{B\pi}^{\,(\mu)} (16~\mathrm{GeV}^2)$ & $2.4(2)\times 10^{-2}$ & $4(2)\times 10^{-2}$ & \cite{Zyla:2020zbs,Lees:2012vv,Sibidanov:2013rkk} \\ 
$ \widehat{r}_{B_s K}^{\,(\mu)} (16~\mathrm{GeV}^2)$ & $1.7(1)\times 10^{-2}$ & -- &  \\ 
$ \widehat{r}_{B\pi}^{\,(\tau)} (16~\mathrm{GeV}^2)$ & $5.4(3)$ & $>0.44$ & \cite{Zyla:2020zbs,Hamer:2015jsa} \\ 
$ \widehat{r}_{B_s K}^{\,(\tau)} (16~\mathrm{GeV}^2)$ & $3.8(2)$ & -- & \\ \hline
\end{tabular}
 \caption{\small \sl Experimental results and our SM predictions for the observables defined in Eq.~\eqref{eq:Rcut} and \eqref{eq:ratio-lep-semilep-Bpi} for $q^2_\mathrm{min} = 16~\mathrm{GeV}^2$.}
\label{tab:obs-BtoPi} 
\end{table}

\noindent where $q^2_\mathrm{min} \geq m_\ell^2$ is to be chosen in auch a way as to avoid the problematic low $q^2$-region. This observable is plotted in Fig.~\ref{fig:Rpi-cut} as a function of $q^2_\mathrm{min}$, where we see that choosing $q^2_\mathrm{min}\gtrsim 10~\mathrm{GeV}^2$ is already enough to obtain consistent results with both approaches. In order to be conservative, we take $q^2_\mathrm{min}=16~\mathrm{GeV}^2$, which also corresponds to one of the $q^2$-bins considered in the experimental measurement of $B\to \pi l \bar{\nu}$ (with $l=e,\mu$) at BaBar~\cite{Lees:2012vv} and Belle~\cite{Sibidanov:2013rkk}. For this choice of integration interval, we obtain the following SM predictions,
\begin{equation}
 \widehat{R}_{B\pi}^{(\tau/\mu)}(16~\mathrm{GeV}^2)\Big{\vert}_\mathrm{LQCD} =1.08(3)\,, \qquad\quad   R_{B\pi}^{(\tau/\mu)}(16~\mathrm{GeV}^2)\Big{\vert}_{\mathrm{LQCD}+\mathrm{exp}} = 1.07(2)\,,
\end{equation}
\noindent which are in perfect agreement. By using the same approach, we define the ratio of semileptonic and leptonic decays as
\begin{align}
\label{eq:ratio-lep-semilep-Bpi}
\widehat{r}_{PP^\prime}^{\,(\ell)}(q^2_\mathrm{min}) \equiv \dfrac{\mathcal{B}(P^{\prime\prime}\to \ell {\nu})}{\displaystyle\int_{q_\mathrm{min}^2}^{(M-m)^2}  \dfrac{\mathrm{d}\overline{\mathcal{B}}}{\mathrm{d}q^2} (P \to P^{\prime} \ell {\nu})\,\mathrm{d}q^2}\,,
\end{align}

\noindent where the denominator accounts for the isospin average from Eq.~\eqref{eq:isospin-average}, and $P^{\prime\prime}$ is defined as in Eq.~\eqref{eq:ratio-lep-semilep}, i.e.~$P^{\prime\prime}=B^+$ for $B\to \pi \ell \bar{\nu}$ and $B_s\to K\ell \bar{\nu}$. Our predictions for these observables are collected in Table~\ref{tab:obs-BtoPi}, along with the existing experimental results. Currently, there is an experimental limit on the decay mode $\mathcal{B}(B\to \pi \tau \bar{\nu})<2.5\times 10^{-4}$~\cite{Hamer:2015jsa}, which is expected to be measured soon at Belle-II with a precision of $\mathcal{O}(20~\%)$~\cite{Kou:2018nap}. For the reasons explained above it would be very useful to separate the low and high-$q^2$ regions. Note also that the ratio of the $B_s\to K$ and $B_s\to D_s$ form factors has been studied in LQCD in Ref.~\cite{Monahan:2018lzv}. The first experimental determination of the ratio of branching fractions of these modes has been reported while this paper was in writing~\cite{Aaij:2020nvo}. In that paper the authors indeed make distinction between the low and high $q^2$ regions, but with $q^2_\mathrm{min}=7\,\gev^2$ that is perhaps too low.

\section{New Physics Phenomenology}
\label{sec:numerics-bis}

In this Section we use the observables discussed in Sec.~\ref{sec:numerics} to constrain the effective couplings defined in Eq.~\eqref{eq:left}, which are then used to explore the new semileptonic observables proposed in Sec.~\ref{sec:semileptonic}. In our analysis, we will focus on the LFU ratios of type $\mu/e$ and $\tau/\mu$, and we will assume that NP couplings affect the decay to the heavier lepton in each ratio (i.e.~$\mu$'s for $\mu/e$ ratios and $\tau$'s for $\tau/\mu$). In other words, our analysis is based on the assumption,
\begin{equation}
\label{eq:flavor-hierarchy}
\qquad\qquad|g_\alpha^{ij\,e}| \ll |g_\alpha^{ij\,\mu}| \ll |g_\alpha^{ij\,\tau}|\,,\qquad \quad\forall~i,j\,
\end{equation}

\noindent which holds true, for instance, in many NP scenarios aiming at explaining the hierarchy of fermion masses, cf. e.g.~Ref.~\cite{Barbieri:2011ci,Fuentes-Martin:2019mun}. However, the theoretical inputs given in Sec.~\ref{sec:numerics} are sufficient to recast our results to a more general NP scenario rather than the one defined in Eq.~\eqref{eq:flavor-hierarchy}. 

The experimental inputs used in our analysis are
\begin{itemize}
 \item[i)] The ratios of semileptonic decays $R^{(\ell/\ell^\prime)}_{PP^\prime}=\mathcal{B}(P\to P^\prime \ell \bar{\nu})/\mathcal{B}(P\to P^\prime \ell^\prime \bar{\nu})$, which are listed in Table~\ref{tab:observables} for the various transitions. 
 \item[ii)] The ratios of leptonic decays $R^{(e/\mu)}_{K}=\mathcal{B}(K\to e\bar{\nu})/\mathcal{B}(K\to \mu\bar{\nu})$ and $R^{(\tau/\mu)}_{K}=\mathcal{B}(\tau\to K{\nu})/\mathcal{B}(K\to \mu\bar{\nu})$, which is given in Table~\ref{tab:observables}.
 \item[iii)] The ratios of leptonic and semileptonic decays $r_{PP^\prime}^{(\ell/\ell^\prime)}\equiv \mathcal{B}(P\to \ell \bar{\nu})/\mathcal{B}(P\to P^\prime \ell^\prime \bar{\nu})$\,,
are simply the products of $R^{(\ell/\ell^\prime)}_{PP^\prime}$ and $r^{(\ell)}_{P P^\prime}$ already presented in Tables~\ref{tab:observables} and~\ref{tab:observables-bis}, respectively. 
\end{itemize}

\noindent Note that for most transitions we opt for using the ratio $r_{PP^\prime}^{(\ell/\ell^\prime)}$, instead of the purely leptonic one, $R_P^{(\ell/\ell^\prime)}=\mathcal{B}(P\to \ell \bar{\nu})/\mathcal{B}(P\to \ell^\prime \bar{\nu})$, since the decays $P\to \ell^\prime \bar{\nu}$ (with $\ell^\prime =e,\mu$) are very rare and still unobserved for many transitions. The only exception is the kaon sector, where $R^{(e/\mu)}_{K}$ and $R^{(\tau/\mu)}_{K}$ have been precisely measured, and in fact used in our analysis~\cite{Zyla:2020zbs}. In addition to the observables listed above, we also consider the ones corresponding to the $B\to \pi \ell \bar{\nu}$, with the choice of the cut $q^2\geq16~\mathrm{GeV}^2$, as described in Sec.~\ref{ssec:BtoPi}.

\subsection{Simplified semileptonic expressions}

Let us discuss the sensitivity of the different semileptonic observables defined in Sec.~\ref{sec:semileptonic} to the NP couplings defined in Eq.~\eqref{eq:left}. Starting from the integrated branching fraction, without loss of generality, we can write \
\begin{align}
\begin{split}
\label{eq:magic-number}
\dfrac{\mathcal{B}_\mathrm{tot}}{\mathcal{B}_\mathrm{tot}^\mathrm{SM}} &= |1+g_{V}|^2 + a_S^{\mathcal{B}}\, |g_S|^2 + a_T^{\mathcal{B}}\, |g_T|^2\\
& + a_{SV}^{\mathcal{B}}\mathrm{Re}\big{[}(1+g_{V})\, g_S^\ast\big{]}+ a_{TV}^{\mathcal{B}}\,\mathrm{Re}\big{[}(1+g_{V})\, g_T^\ast\big{]}+ a_{ST}^{\mathcal{B}}\,\mathrm{Re}\big{[}g_S\, g_T^\ast\big{]}\,,
\end{split}
\end{align}

\begin{table}[t]
\renewcommand{\arraystretch}{1.8}
\centering
\begin{tabular}{|c|c|ccccc|}
\hline 
Decay  & $|V_{ij}|^{-2}\,\overline{\mathcal{B}}(P\to P^\prime\ell\bar{\nu})$ & $a_S^{\mathcal{B}}$ & $a_T^{\mathcal{B}}$ & $a_{SV_L}^{\mathcal{B}}$ & $a_{TV_L}^{\mathcal{B}}$ & $a_{ST}^{\mathcal{B}}$ \\ \hline\hline
$K^+\to \pi^0 \mu \bar{\nu}$ &  $0.669(6)$ & $15.74(12)$ & $0.152(11)$ &  $4.43(3)$ & $0.46(2)$& $0$ \\ 
$D^+\to \pi^0 \mu \bar{\nu}$ & $0.066(4)$ &  $2.39(12)$ & $1.17(18)$ &  $0.435(15)$ & $0.47(4)$ & $0$ \\ 
$D^+\to \overline{K^0} \mu \bar{\nu}$  &$0.091(6)$ & $1.69(5)$ &$0.71(10)$ & $0.465(10)$ &$0.45(3)$ & $0$ \\ 
$B^+\to D^0 \mu \bar{\nu}$  &$14.8(8)$ & $1.13(3)$  &  $0.68(6)$ & $0.154(2)$ & $0.188(9)$ & $0$ \\
$B^+\to D^0 \tau \bar{\nu}$ & $4.3(1)$ & $1.076(9)$ & $0.84(8)$ & $1.533(9)$ & $1.09(5)$ & $0$ \\ \hline
      \end{tabular}
\caption{\small \sl Numerical coefficients entering Eq.~\eqref{eq:magic-number} for the different semileptonic transitions. We also quote the values for the SM predictions $\mathcal{B}_\mathrm{tot}^\mathrm{SM}=\mathcal{B}(P\to P^\prime\ell\bar{\nu})$ after factoring out the CKM matrix elements $|V_{ij}|$. As mentioned in the text, the renormalization scale for all the coefficients is taken to be $\mu =2\,\gev$, except for the $B$-meson decays for which $\mu = m_b$. }
\label{tab:magic-numbers} 
\end{table}

\noindent  where $a^{\mathcal{B}}_\alpha$ are the numerically known coefficients obtained by integrating over the full range of $q^2$'s. Note that the flavor indices in $g_\alpha \equiv g_\alpha^{ij\,\ell}$ are omitted. We evaluated all of $a^{\mathcal{B}}_\alpha$ and collected the results in Table~\ref{tab:magic-numbers} for each of the transitions considered in this paper. These values can be combined with the SM predictions quoted in Table~\ref{tab:observables} to compute the LFU ratios defined in Eq.~\eqref{eq:ratio-lep} for the most general NP scenario. For the $B\to\pi\ell \bar{\nu}$ transition, we list the coefficients $a^{\mathcal{B}}_\alpha\equiv a^{\mathcal{B}}_\alpha(q^2_\mathrm{min})$ in Table~\ref{tab:magic-numbers-Bpi}, as obtained for different values of $q^2_\mathrm{min}$ and by using the LQCD form factors from Refs.~\cite{Lattice:2015tia,Flynn:2015mha}. Notice that the coefficient $a_{ST}^{\mathcal{B}}$ vanishes identically. This particular combination of effective couplings $\propto g_S g_T^\ast$  can only be probed by using the full angular distribution, as we discuss in the following.

For the semileptonic observables $\mathcal{O}\in \lbrace A_\mathrm{fb},~A_\lambda,~A_{\pi/3}\rbrace$ defined in Sec.~\ref{sec:semileptonic}, we can write in full generality,
\begin{align}
\begin{split}
\label{eq:magic-number-bis}
\langle \mathcal{O}\rangle \, \dfrac{\mathcal{B}_\mathrm{tot}}{\mathcal{B}_\mathrm{tot}^\mathrm{SM}} &= {\langle \mathcal{O}^\mathrm{SM}\rangle}\,|1+g_{V}|^2 + b_S^{\mathcal{O}}\, |g_S|^2 + b_T^{\mathcal{O}}\, |g_T|^2\\[0.4em]
& + b_{SV}^{\mathcal{O}}\mathrm{Re}\big{[}(1+g_{V})\, g_S^\ast\big{]}+ b_{TV}^{\mathcal{O}}\,\mathrm{Re}\big{[}(1+g_{V})\, g_T^\ast\big{]}+ b_{ST}^{\mathcal{O}}\,\mathrm{Re}\big{[}g_S\, g_T^\ast\big{]}\,,
\end{split}
\end{align}
where $\mathcal{B}_\mathrm{tot} \equiv \mathcal{B}_\mathrm{tot}(g_V,g_S,g_T)$ is the total branching fraction, $b_\alpha^\mathcal{O}$ are the known numerical coefficients, and the brackets $\langle \dots \rangle$ denote the integration over the full $q^2$ range,~\footnote{In this notation the total branching fraction can be written as $B_\mathrm{tot}= \big{\langle} \mathcal{B}(P\to P^\prime \ell \bar{\nu})\big{\rangle}$.}
\begin{align}
\label{eq:int-obs}
\langle \mathcal{O}\rangle = \int_{m_\ell^2}^{(M-m)^2} \dfrac{\mathrm{d}\mathcal{O}}{\mathrm{d}q^2} \,\mathrm{d}q^2 \,.
\end{align}
The values of all coefficients $b_\alpha^\mathcal{O}$ are collected in Table~\ref{tab:magic-numbers-newobs}. By comparing Table~\ref{tab:magic-numbers} and \ref{tab:magic-numbers-newobs}, it is evident that $A_{\mathrm{fb}}$, $A_{\lambda}$ and $A_{\pi/3}$ are complementary to the branching fractions. In particular, $A_\mathrm{fb}$ is the only observable that depends on $\mathrm{Re}(g_S\, g_T^\ast)$, with an enhanced sensitivity due to a large numerical coefficients $b_{ST}^{A_\mathrm{fb}}$, cf. Table~\ref{tab:magic-numbers-newobs}. To assess the potential of these new observables to reveal the presence of NP, we first need to determine the allowed ranges of the effective NP couplings entering Eq.~\eqref{eq:magic-number-bis}.

\begin{table}[t]
\renewcommand{\arraystretch}{1.8}
\centering
\begin{tabular}{|c|c|c|ccccc|}
\hline 
Decay  & $q^2_\mathrm{min}$ & $|V_{ub}|^{-2}\,B_\mathrm{tot}(q^2\geq {q^2_\mathrm{min}})$ & $a_S^{\mathcal{B}}$ & $a_T^{\mathcal{B}}$ & $a_{SV_L}^{\mathcal{B}}$ & $a_{TV_L}^{\mathcal{B}}$ & $a_{ST}^{\mathcal{B}}$ \\ \hline\hline
\multirow{3}{*}{$B^+\to \pi^0 \mu \bar{\nu}$} & $12~\mathrm{GeV}^2$ & $2.1(2)$ & $2.8(3)$ & $5(1)$ & $0.13(1)$ & $0.23(3)$ & $0$ \\ 
& $16~\mathrm{GeV}^2$ & $1.4(1)$ & $3.6(3)$ & $5.2(6)$ & $0.15(1)$ & $0.23(1)$ & $0$ \\ 
 & $20~\mathrm{GeV}^2$ & $0.66(3)$ & $5.2(4)$ & $5.8(5)$ & $0.20(1)$& $0.23(1)$ & $0$ \\ \hline
\multirow{3}{*}{$B^+\to \pi^0 \tau \bar{\nu}$} &  $12~\mathrm{GeV}^2$ & $2.2(2)$ & $2.0(1)$ & $4.3(9)$ & $1.5(1)$ & $2.6(3)$ & $0$ \\
 &  $16~\mathrm{GeV}^2$ & $1.5(1)$ & $2.4(1)$ & $4.5(5)$ & $1.7(1)$ & $2.5(1) $& $0$ \\ 
 &  $20~\mathrm{GeV}^2$ & $0.78(3)$ & $3.3(1)$ & $4.6(4)$ & $2.1(1)$ & $2.4(1)$ & $0$ \\ \hline
      \end{tabular}
\caption{\small \sl Numerical coefficients $a_\alpha^{\mathcal{B}}\equiv a_\alpha^{\mathcal{B}}(q^2_\mathrm{min})$ appearing in Eq.~\eqref{eq:magic-number} for the decays $B\to \pi \ell \bar{\nu}$ in the interval $q^2 \in (q^2_\mathrm{min},(m_B-m_\pi)^2)$ with $q^2_\mathrm{min}$ fixed.}
\label{tab:magic-numbers-Bpi} 
\end{table}

\begin{table}[p]
\renewcommand{\arraystretch}{1.9}
\centering
\begin{tabular}{|c|c|c|ccccc|}
\hline 
Decay mode & $\mathcal{O}$ & $\big{\langle}\mathcal{O}^\mathrm{SM}\big{\rangle}$ & $b_S^{\mathcal{O}}$ & $b_T^{\mathcal{O}}$ & $b_{SV_L}^{\mathcal{O}}$ & $b_{TV_L}^{\mathcal{O}}$ & $b_{ST}^{\mathcal{O}}$ \\ \hline\hline
\multirow{3}{*}{$K^-\to \pi^0 \mu \bar{\nu}$}  & $A_\mathrm{fb}$ & $0.2726(3)$ & $0$ & $0$  & $1.379(2)$ & $0.343(13)$ & $2.15(8)$\\
 & $A_{\pi/3}$ & $-0.1537(6)$ & $0$ & $0.066(5)$ & $0$& $0$& $0$\\
 & $A_{\lambda}$ & $-0.091(4)$ & $15.79(11)$ & $0.065(4)$ & $4.43(3)$ & $-0.154(6)$& $0$ \\ \hline
\multirow{3}{*}{$D^-\to \pi^0 \mu \bar{\nu}$}  & $A_\mathrm{fb}$ & $0.0386(11)$ & $0$ & $0$ & $0.160(2)$ & $0.29(3)$ & $2.7(2)$\\
 & $A_{\pi/3}$ & $-0.3455(8)$ & $0$ & $0.84(13)$ & $0$& $0$& $0$\\
 & $A_{\lambda}$ & $-0.890(3)$ & $2.40(12)$ & $1.1(2)$ & $0.435(14)$ & $-0.156(14)$ & $0$ \\ \hline
\multirow{3}{*}{$D^-\to K^0 \mu \bar{\nu}$}  & $A_\mathrm{fb}$ & $0.0580(8)$ & $0$ & $0$ & $0.1714(15)$ & $0.29(2)$ & $1.78(12)$\\
 & $A_{\pi/3}$ & $-0.3307(7)$ & $0$ & $0.51(7)$ & $0$& $0$& $0$\\
 & $A_{\lambda}$ & $-0.833(3)$ & $1.69(5)$ & $0.66(9)$ & $0.465(10)$ & $-0.150(10)$ & $0$ \\ \hline
\multirow{3}{*}{$B^-\to D^0 \mu \bar{\nu}$}  & $A_\mathrm{fb}$ & $0.0141(3)$& $0$  & $0$ & $0.0590(4)$ & $0.116(5)$ & $1.45(7)$\\
 & $A_{\pi/3}$ & $-0.3643(2)$ & $0$ & $0.50(5)$ & $0$& $0$& $0$\\
 & $A_{\lambda}$ & $-0.9605(8)$ & $1.13(3)$ & $0.67(6)$ & $0.154(2)$ & $-0.062(3)$ & $0$ \\ \hline
\multirow{3}{*}{$B^-\to D^0 \tau \bar{\nu}$}  & $A_\mathrm{fb}$ & $0.3602(8)$ & $0$ & $0$ & $0.4430(8)$ & $0.87(4)$ & $1.14(5)$ \\
 & $A_{\pi/3}$ & $-0.0671(3)$ & $0$ & $0.18(2)$ & $0$& $0$& $0$\\
 & $A_{\lambda}$ & $0.324(3)$ & $1.076(10)$ & $0.052(5)$ & $1.534(10)$ & $-0.36(2)$ & $0$ \\ \hline
 \end{tabular}
\caption{\small \sl Numerical coefficients for the coefficients $b_i^\mathcal{O}$  defined in Eq.~\eqref{eq:magic-number-bis} for the integrated observables $\mathcal{O}\in \lbrace A_\mathrm{fb}, A_{\pi/3}, A_\lambda \rbrace$ defined in Sec.~\ref{sec:semileptonic}. Notice that the ``magic numbers" are given for the decays of charged mesons, but that they are practically if one considers decays of neutral mesons for the quantities as defined in Eq.~\eqref{eq:magic-number-bis}. }
\label{tab:magic-numbers-newobs} 
\end{table}

\subsection{Constraints and predictions}

To determine the allowed ranges of the NP effective couplings we consider the observables described above, with the experimental results and SM predictions given in Tables~\ref{tab:observables}, \ref{tab:observables-bis} and \ref{tab:obs-BtoPi}. In addition to these observables, we also require that $\mathcal{B}(B_c\to \ell \bar{\nu})\lesssim 30~\%$ in order to avoid the saturation of the $B_c$-meson lifetime, the value of which is known experimentally~\cite{Alonso:2016oyd}.These quantities are used in Table~\ref{tab:constraints} to constrain the couplings $g_A^{ij\,\ell}$ and $g_P^{ij\,\ell}$ from the leptonic decays, and $g_V^{ij\,\ell}$, $g_S^{ij\,\ell}$ and $g_T^{ij\,\ell}$ from the semileptonic ones. The renormalization scale $\mu$ is taken to be $\mu=2~\mathrm{GeV}$ for the decays of $K$ and $D$-mesons, and $\mu=m_b$ for $b$-decays. Several comments regarding the results are in order:

\begin{itemize}
 \item First, we note that there are two distinct real solutions for each NP coupling due to the quadratic dependence of the branching fraction on $g_{\alpha}^{ij\,\ell}$, as it can be seen in Eqs.~\eqref{eq:magic-number}. In Table~\ref{tab:constraints}, we choose the solution closer to the SM, since the other one would correspond to a NP scenarios with large NP couplings which is most likely in tension with the direct searches at LHC. 
 
  \item Our analysis was based on the assumption that the NP couplings to leptons are hierarchical, see Eq.~\eqref{eq:flavor-hierarchy}. The CKM matrix element is eliminated in the ratios of leptonic or semileptonic decays differing in flavor of the lepton in the final state. 
 
 \item For the semileptonic decays based on the transitions $s\to u \tau \nu$, $c\to d \tau \nu$ and $c\to s \tau \nu$, there is no available phase space which is why the corresponding $g_{V}^{ij\,\tau}$, $g_S^{ij\,\tau}$ and $g_T^{ij\,\tau}$ effective couplings are not constrained by the low-energy data. 
 
 \item The decays $B\to \pi \ell \bar{\nu}$ with $\ell=e,\mu$ are systematically combined in the experimental analyses performed at the $B$-factories~\cite{Lees:2012vv,Sibidanov:2013rkk}. While this is the best approach to extracting the $|V_{ub}|$ value, it is not straightforward to use these results in order to constrain the NP scenarios in which the LFU is broken, as we assume. For this reason, we prefer not to quote any constraint for this particular transition. We suggest to the future experimental analyses to also quote the value of $R_{B\pi}^{(\mu/e)} = \mathcal{B}(B\to \pi \mu \bar{\nu})/\mathcal{B}(B\to \pi e \bar{\nu})$, as done for instance in certain studies of $B\to D \ell \bar{\nu}$ decays~\cite{Glattauer:2015teq}.
 
  \item The only significant discrepancy between theory and experiment in Table~\ref{tab:constraints} is the well-known $B$-physics LFU deviation in the $B\to D\ell \bar{\nu}$ transition~\cite{Lees:2012xj,Huschle:2015rga}. For this particular transition, the allowed range for the effective couplings would become more constrained if results concerning the $B\to D^\ast \tau \nu$ transition were also considered, see e.g.~Ref.~\cite{Becirevic:2019tpx}. Note also that the small deviations observed in $D^0\to \pi^+\mu \bar{\nu}$ decays become less significant when the isospin average is considered, as discussed in Sec.~\ref{sec:numerics}.  
 \end{itemize}

\begin{table}[t]
\centering
\resizebox{\columnwidth}{!}{%
\renewcommand{\arraystretch}{1.95}
\begin{tabular}{|c|ccccc|}
\hline 
$u_i \,d_j\,\ell$ & $\mathrm{Re}\big{(}g_V^{ij\,\ell}\big{)}$ & $\mathrm{Re}\big{(}g_A^{ij\,\ell}\big{)}$ & $\mathrm{Re}\big{(}g_S^{ij\,\ell}\big{)}$& $\mathrm{Re}\big{(}g_P^{ij\,\ell}\big{)}$ & $\mathrm{Re}\big{(}g_T^{ij\,\ell}\big{)}$\\ \hline\hline
$u\,s\,\mu$ & $(0 \pm 2)\times 10^{-3}$  &$(2.2\pm 1.8) \times 10^{-3}$ & $(-2\pm 9)\times 10^{-4}$ &  $(-9\pm 8) \times 10^{-5}$ & $(-2\pm 9)\times 10^{-3}$ \\ 
$u\,s\,\tau$ & -- & $(2.2\pm 1.7)\times 10^{-2}$ & -- & $(1.6\pm 1.1)\times 10^{-2}$ & -- \\ 
$c\,d\,\mu$ & $(-3.0\pm 1.6)\times 10^{-2}$ & $(7\pm 4)\times 10^{-2}$ & $(-9\pm 7)\times 10^{-2}$ & $(-2.6\pm 1.3)\times 10^{-3}$ & $(-2.0\pm 1.4)\times 10^{-1}$\\ 
$c\,d\,\tau$ & -- & $(-0.1\pm 1.1)\times 10^{-1}$ & -- & $(1\pm 7)\times 10^{-2}$ & -- \\
$c\,s\,\mu$ & $(3\pm 6)\times 10^{-3}$ & $(-2\pm 4)\times 10^{-2}$ & $(-1\pm 2)\times 10^{-2}$ & $(0.7\pm 1.4)\times 10^{-3}$ & $(1.2 \pm 2.7)\times 10^{-2}$ \\  
$c\,s\,\tau$ & -- & $(-3\pm 4)\times 10^{-2}$ & -- & $(2\pm 2)\times 10^{-2}$ & --  \\
$u\,b\,\mu$ & -- & --& -- & -- & --\\  
$u\,b\,\tau$ & $-1\pm 2$ & $(-1 \pm 2)\times 10^{-1}$ & $-0.3\pm 1.5$ & $(3 \pm 7)\times 10^{-2}$ & $-0.3\pm 1.1$ \\
$c\,b\,\mu$ & $(0\pm 2)\times 10^{-2}$ & -- & $(1\pm 2)\times 10^{-1}$ & $(0\pm 8 )\times 10^{-1}$ &  $(-1\pm 3)\times 10^{-1}$ \\ 
$c\,b\,\tau$ & $(7\pm 5)\times 10^{-2}$ & $1\pm 4$ & $(9\pm 6)\times 10^{-2}$ &  $(-2\pm 8)\times 10^{-1}$ & $(1.2\pm 0.8)\times 10^{-1}$  \\ \hline
\end{tabular} 
}
\caption{\small \sl $1\sigma$ constraints on the real part of the coefficients $g_\alpha^{ij\,\ell} = g_\alpha^{ij\,\ell}(\mu)$, with $\alpha\in \lbrace V,A,S,P,T \rbrace$), derived from the observables collected in Table~\ref{tab:observables}. The scale $\mu$ is taken to be $\mu=2~\mathrm{GeV}$ for $K$ and $D$-meson observables, and $\mu=m_b$ for $B$-meson decays.}
\label{tab:constraints} 
\end{table}

\begin{figure}[p!]
\centering
\vspace*{-4.em}
\includegraphics[width=.94\linewidth]{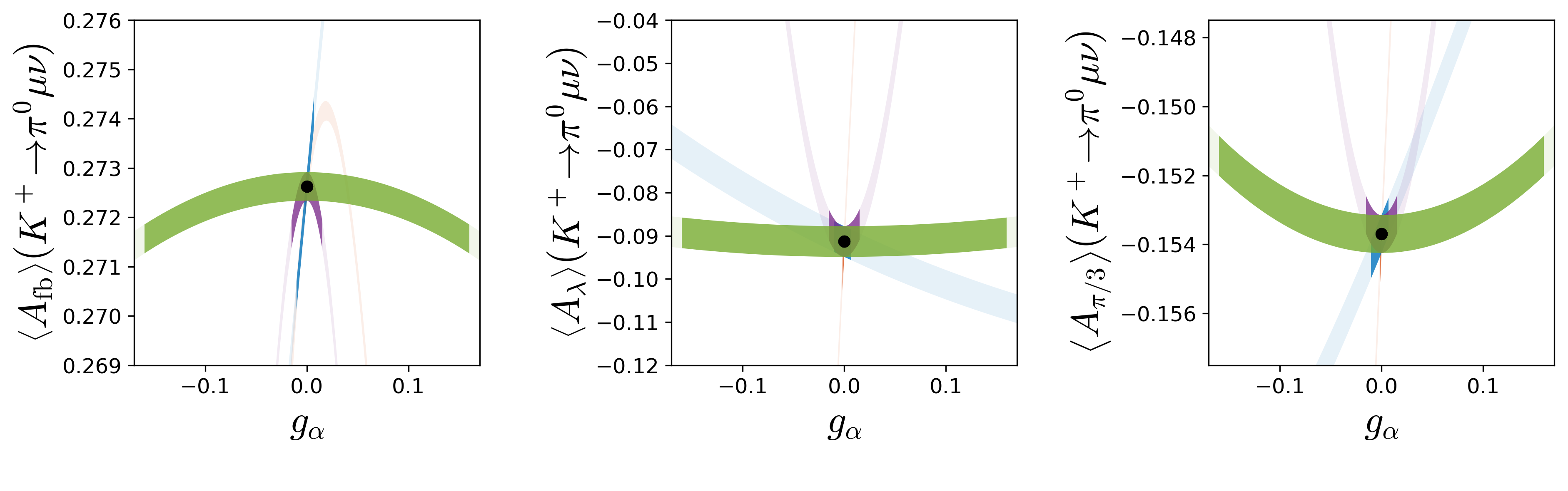}\\
\vspace*{-1.2em}
\includegraphics[width=.94\linewidth]{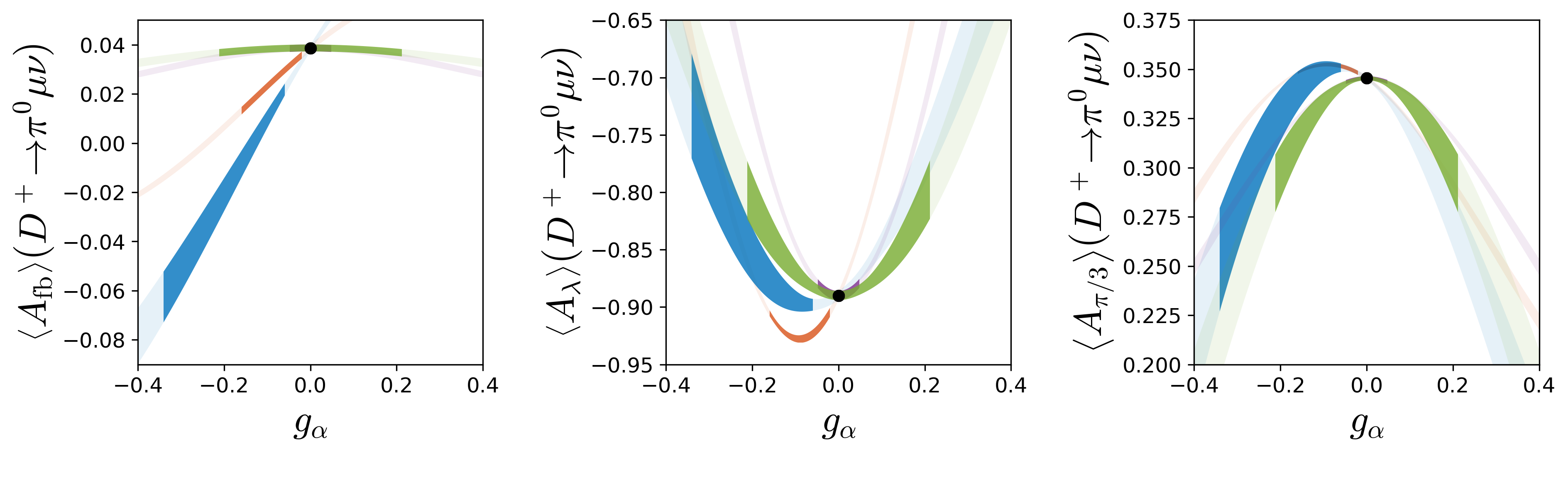}\\
\vspace*{-1.2em}
\includegraphics[width=.94\linewidth]{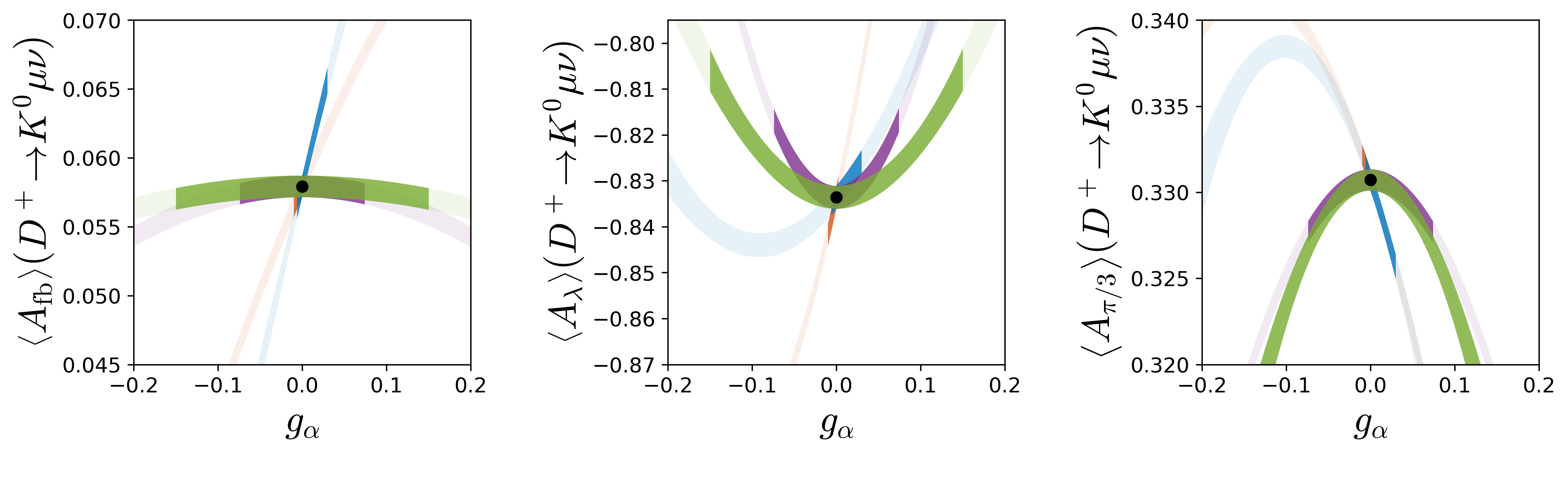}\\
\vspace*{-1.2em}
\includegraphics[width=.94\linewidth]{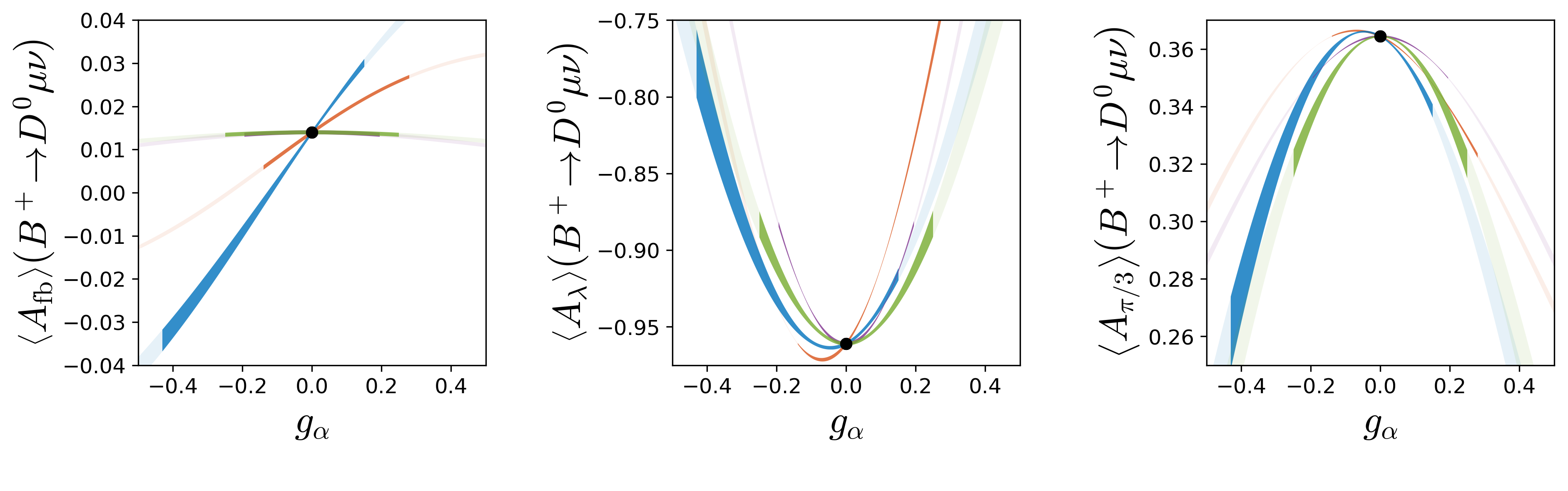}\\
\vspace*{-1.2em}
\includegraphics[width=.94\linewidth]{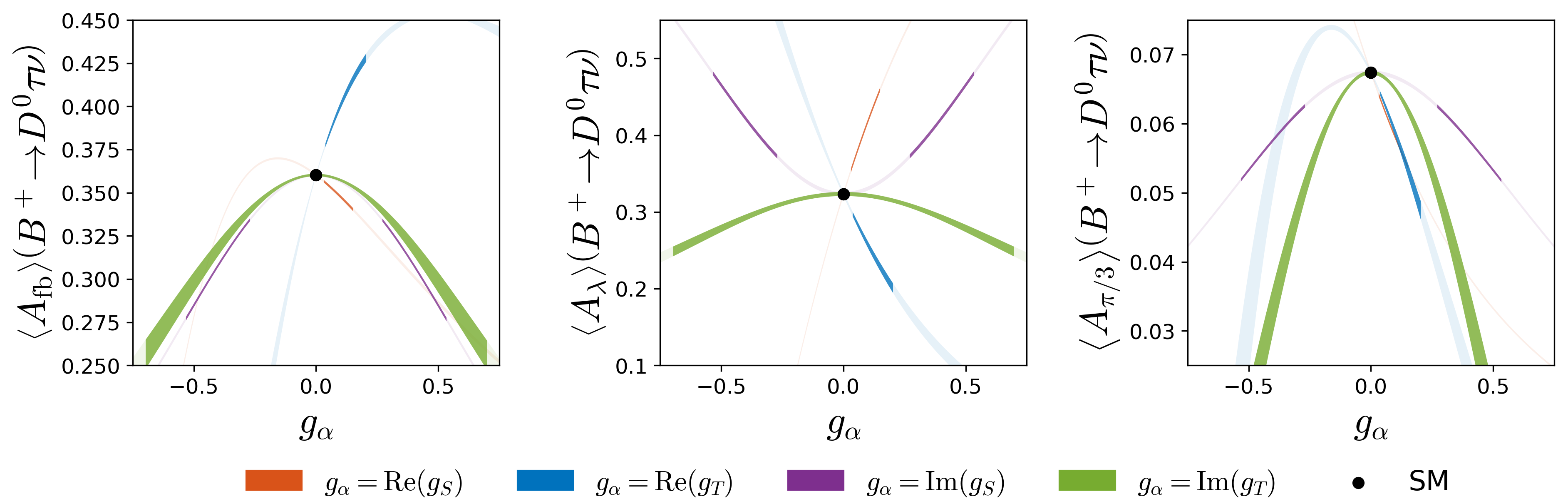}\\[.3em]
\caption{\small \sl Predictions for the integrated observables $\langle A_\mathrm{fb}\rangle$, $\langle A_\lambda\rangle$ and $\langle A_{\pi/3}\rangle$, defined in Eq.~\eqref{eq:int-obs}, as a function of the Wilson coefficients $g_i\in \lbrace\mathrm{Re}(g_{S}), \mathrm{Re}(g_{T}), \mathrm{Im}(g_{S}), \mathrm{Im}(g_{T}) \rbrace$. The darker regions are allowed by existing experimental constraints collected in Table~\ref{tab:observables}. } 
\label{fig:new-obs}
\end{figure}

We are now in a position to use the constraints obtained in Table~\ref{tab:constraints} and predict the value of new observables $A_{\mathrm{fb}}$, $A_{\lambda}$ and $A_{\pi/3}$, defined in Sec.~\ref{sec:semileptonic}, as a function of the allowed ranges for the NP couplings. We first discuss their integrated values, see Eq.~\eqref{eq:int-obs}. These quantities are plotted in Fig.~\ref{fig:new-obs} as functions of the real and imaginary parts of $g_{S}^{ij\,\alpha}$ and $g_{T}^{ij\,\alpha}$, for each quark-level transition. The light colored regions show the dependence of the physical observables on the effective NP couplings, whereas the values allowed by the constraints given in Table~\ref{tab:constraints} are highlighted by darker colors. In that plot, we see that the sizeable deviations from the SM are indeed possible. For instance, $\langle A_\mathrm{fb}\rangle(D\to \pi \mu\bar{\nu})$ can be modified by varying the NP coupling $g_T$ in the interval allowed by the data. Its value could not only change the sign but its absolute value could be $\approx 2\times$ larger than its SM value. Significant deviations for $A_\mathrm{fb}$ and $A_{\pi/3}$, are also possible in $D\to K \mu \bar{\nu}$, $B\to D \mu \bar{\nu}$ and $B\to D \tau \bar{\nu}$. It is therefore clear that studying the angular distribution of these decays experimentally could offer a fertile ground for searching the NP effects.

For decays to $\tau$-leptons, such as $B_{(s)}\to D_{(s)}\tau \bar{\nu}$, $B_{s}\to K\tau \bar{\nu}$ and $B\to \pi\tau \bar{\nu}$, the $\tau$-polarization is also experimentally accessible, since it can be reconstructed from the kinematics of its decay products~\cite{Asadi:2020fdo,Alonso:2017ktd}. From Fig.~\ref{fig:new-obs} we see that the lepton-polarization asymmetry $\langle A_\lambda \rangle(B\to D \tau \bar{\nu})$ is very sensitive to the NP couplings, which can be increased (decreased) by a pronounced NP coupling to the scalar (tensor) operator. For the processes involving muons, it is not clear how the lepton polarization can be determined since muons are stable for the length scales probed in most particle colliders. For these decays, the only observables that can be reconstructed with known techniques are $A_\mathrm{fb}$ and $A_\mathrm{\pi/3}$, and the predictions for $A_\lambda$ are less relevant, being given in Fig.~\ref{fig:new-obs} only for the sake of completeness.

\begin{figure}[t!]
\centering
\includegraphics[width=1.\linewidth]{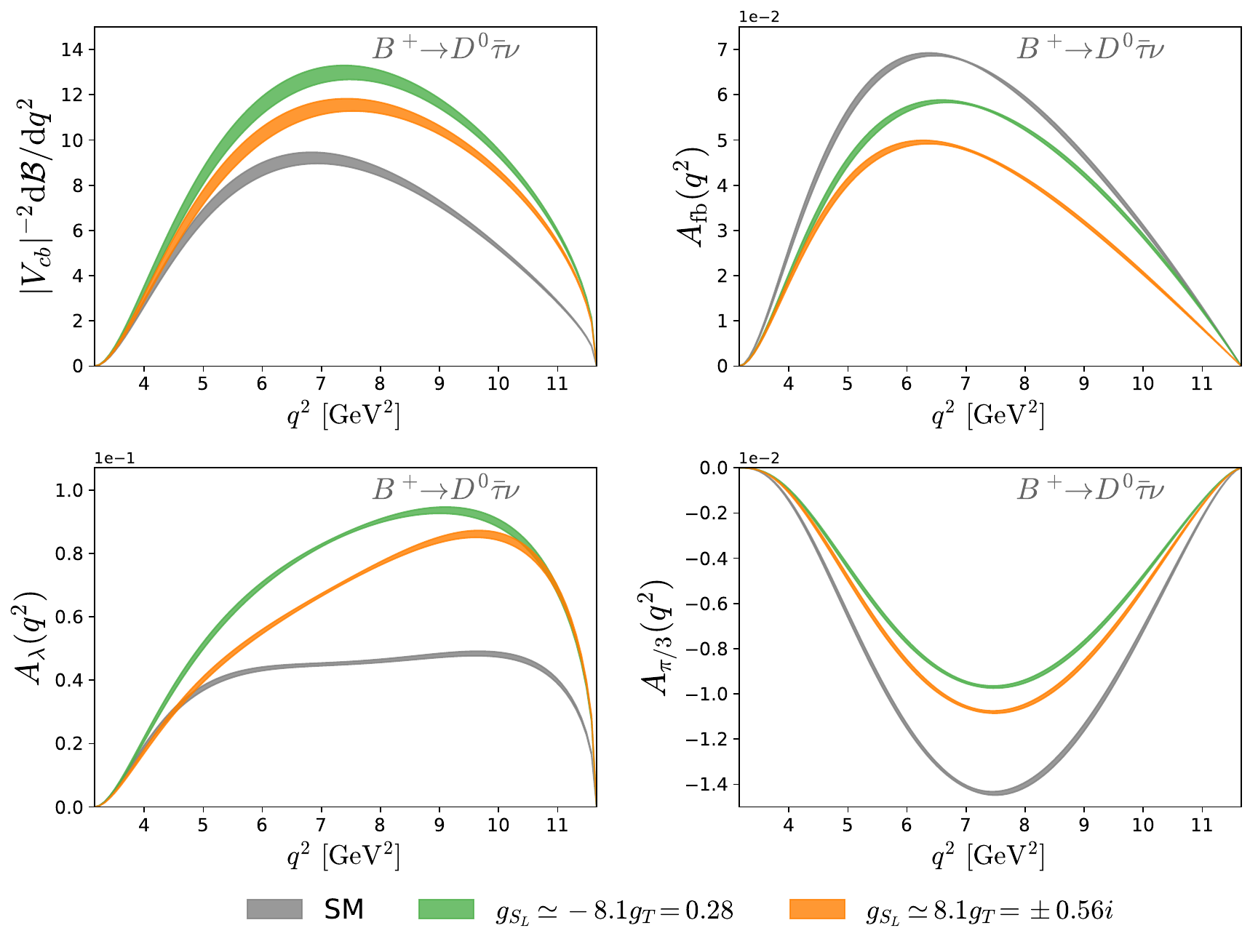}
\caption{\small \sl Predictions for the differential distributions of $A_\mathrm{fb}(q^2)$, $A_\lambda(q^2)$ and $A_{\pi/3}(q^2)$ for the $B\to D \tau \bar{\nu}$ transition. The benchmark values for the NP scenarios are motivated by the LQ scenarios that can accommodate the discrepancies observed in $B\to D^{(\ast)}l\bar{\nu}$~\cite{Angelescu:2018tyl}. See text for details.}
\label{fig:BDtaunu-differential} 
\end{figure}

Finally, we also explore the impact of NP effects on the differential distributions of the quantities (observables) discussed above. We focus on $B\to D\tau \bar{\nu}$, as motivated by the discrepancies observed in $B$-meson decays~\cite{Lees:2012xj,Huschle:2015rga,Hirose:2016wfn,Aaij:2015yra}. For simplicity, we consider the scenarios in which the SM is extended by a $\mathcal{O}(1\,\tev)$ leptoquark boson $S_1=(\mathbf{\bar{3}},\mathbf{1},1/3)$ or $R_2=(\mathbf{3},\mathbf{2},7/6)$, where in the parentheses are the SM quantum numbers. These scenarios can accommodate the observed LFU discrepancies and remain consistent with numerous low and high-energy constraints~\cite{Angelescu:2018tyl}.~\footnote{Another viable solution to the problem of $B$-anomalies is given by the vector LQ $U_1 = (\bar{3},1,2/3)$, see e.g.~Ref.~\cite{Angelescu:2018tyl} and references therein. Even though this scenario can also allow for a nonzero of $g_{S_R}^{cb\,\tau}$, the dominant coupling to explain the anomalies is $g_{V_L}^{cb\,\tau}$ which does not affect the asymmetries considered above~\cite{Angelescu:2018tyl}.} Moreover, in these models the NP couplings satisfy $g_{S_L}(\Lambda)=-4\, g_T(\Lambda)$ and $g_{S_L}(\Lambda)=+4\, g_T(\Lambda)$, respectively, at the matching scale $\Lambda$. After accounting for the running effects from $\Lambda \approx 1~\mathrm{TeV}$ down to $\mu_b=m_b$, these relations become $g_{S_L}(\mu_b) \approx -8.5\, g_T(\mu_b)$ and $g_{S_L}(\mu_b) \approx 8.14\, g_T(\mu_b)$, respectively. We use the best-fit values for the NP couplings obtained in Ref.~\cite{Angelescu:2018tyl} for these two leptoquark scenarios and plot the differential $q^2$-distributions of different observables. Notice that these values for the effective couplings are determined by using $R_{D^{(\ast)}}^\mathrm{exp}$ which have been extracted experimentally by assuming only the SM for the decay distributions and acceptances and which might also be affected by the NP couplings~\cite{Bernlochner:2020tfi}. The results are shown in Fig.~\ref{fig:BDtaunu-differential}. We find that the overall normalization of $A_\mathrm{fb}$ and $A_\lambda$, as well as the branching fraction, can change by about $ 20\%$ ($S_1$) and by about $50~\%$ ($R_2$), which are possibly large enough to be testable at the LHCb and Belle-II.
Even more significant are the predictions for $A_{\pi/3}$, which can be strongly modified by the plausible values of the NP couplings, especially in the region of intermediate $q^2$'s.

Therefore, measuring the observables discussed in this paper and their $q^2$ shapes can indeed be revelatory of the non-zero value of one of the NP couplings.

\section{Conclusion}
\label{sec:summary}
In this work we made a comprehensive phenomenological analysis of the leptonic and semileptonic decays of pseudoscalar mesons in the framework of a general low energy effective theory which includes all possible interactions BSM, except for possible contributions arising from the right handed neutrinos.

 One of our main goals was to derive the constraints on the NP couplings by relying only on the decay modes for which the non-perturbative QCD uncertainties are fully under control, i.e. which are handled by means of extensive numerical simulations of QCD on the lattice. Such channels are only  those that involve pseudoscalar mesons. By switching on the NP couplings, one at the time, we were able to derive constraints by comparing the accurate theoretical determination with the experimentally available results for the (partial) branching fractions. To eliminate the dependence on the CKM matrix elements we combined similar decay channels in suitable ratios. 
 
The obtained constraints on the NP couplings are then used to predict the possible departure of the angular observables with respect to their SM values. To that effect we showed that one can construct at most four independent observables from the detailed study of the angular distribution of the semileptonic pseudscalar-to-pseudoscalar meson decays. Our results show that these observables can indeed reveal the presence of physics BSM both through their values integrated over the available phase space, or through modification of their $q^2$-dependence with respect to the SM. Clearly more experimental work in this direction is very much needed. 

Besides turning one NP coupling at the time, we also discussed a possibility of simultaneously including two non-zero couplings. Such a situation is realized in the scenarios in which the SM is extended by a low energy scalar leptoquark, such as $R_2$ or $S_1$, for which the scalar and tensor couplings are both nonzero but the ratio of the two being fixed.   

The future analyses along the one presented in this paper should be updated and extended to include the decays to vector mesons in the final state, as long as the vector meson is sufficiently narrow. For that to be done one also needs reliable LQCD results for the form factors, obtained by more than one LQCD collaboration. If these results were
available, we would end up with far more restrictive constraints on the New Physics couplings and many more observables to predict. With the further improvement in accuracy of the experimental results and of the hadronic matrix elements, one also has to start accounting for the electromagnetic corrections. Such a situation is already present in the case of the kaon leptonic and semileptonic decays for which we included electromagnetic corrections as estimated by means of chiral perturbation theory with the low energy constants fixed from phenomenology. The strategies to control the electromagnetic corrections through LQCD studies exist and the first results for the leptonic decays of kaon appeared very recently in Ref.~\cite{Frezzotti:2020bfa} and the result is compatible with what we used in this paper.

\section*{Acknowledgements}

We warmly thank La\'is Sarem Schunk for collaboration at the early stages of this project. This project has received funding from the European Research Council (ERC) under the European Union's Horizon 2020 research and innovation programme under grant agreement 833280 (FLAY), the Marie Skłodowska-Curie grant agreement H2020-MSCA-ITN-2019//860881-HIDDeN, and the Swiss National Science Foundation (SNF) under contract 200021-175940. The research of A.P.M.~was supported by the Cluster of Excellence {\em Precision Physics, Fundamental Interactions and Structure of Matter\/} (PRISMA+ -- EXC~2118/1) within the German Excellence Strategy (project ID 39083149) and by  the BMBF-Project 05H2018 -- Belle II.

\newpage  

\appendix


\section{Matching to the SMEFT}
\label{app:smeft}

 Under the general assumption that NP arises well above the electroweak scale, one should replace Eq.~\eqref{eq:left} by an EFT that is also invariant under $SU(2)_L\times U(1)_Y$, i.e.~the SMEFT~\cite{Grzadkowski:2010es,Buchmuller:1985jz}. The SMEFT Lagrangian can be parameterized as

\begin{equation}
    \label{eq:smeft}
    \mathcal{L}_\mathrm{SMEFT} = \sum_\alpha \dfrac{C_\alpha}{\Lambda^2}\,\mathcal{O}_\alpha\,,
\end{equation}

\noindent where $\Lambda$ is the EFT cutoff, and $C_\alpha$ stand for the effective coefficients of the dimension-6 operators $\mathcal{O}_\alpha$. Only five of these operators can generate  at tree-level the operators in Eq.~\eqref{eq:left}, as listed in Table~\ref{tab:smeft}. In order to match Eq.~\eqref{eq:left} to \eqref{eq:smeft}, we assume that down-quark and lepton Yukawa couplings are diagonal, and that right-handed fermions are in the mass basis. The matching relations at $\mu = \mu_{\mathrm{EW}}$ are then given by
\begin{align}
    \label{eq:smeft-matching}
    g_{V_L}^{ij\,\ell}(\mu_\mathrm{EW}) &= -\dfrac{v^2}{\Lambda^2}\sum_k \dfrac{V_{ik}}{V_{ij}}\Big{(}\Big{[}C_{\substack{lq}}^{(3)}\Big{]}_{\ell\ell kj} +\Big{[}C_{\substack{Hq}}^{(3)}\Big{]}_{kj}-\delta_{kj}\, \Big{[}C_{\substack{Hl}}^{(3)}\Big{]}_{\ell\ell}\,\bigg{)}\,,\nonumber\\[0.35em]
    g_{V_R}^{ij\,\ell}(\mu_\mathrm{EW}) &= 
    \dfrac{v^2}{2\Lambda^2}\dfrac{1}{V_{ij}}\Big{[}C_{{Hud}}\Big{]}_{ij}\,, \nonumber\\[0.35em]
    g_{S_L}^{ij\,\ell}(\mu_\mathrm{EW}) &= -\dfrac{v^2}{2\Lambda^2} \dfrac{1}{V_{ij}} \Big{[}C_{\substack{lequ}}^{(1)}\Big{]}_{\ell\ell ji}^\ast\,,\\[0.35em]
    g_{S_R}^{ij}(\mu_\mathrm{EW}) &= -\dfrac{v^2}{2\Lambda^2} \sum_k\dfrac{V_{ik}^\ast}{V_{ij}}\Big{[}C_{\substack{ledq}}\Big{]}^{\ast}_{\ell\ell jk}\,,\nonumber\\[0.35em]
    g_{T}^{ij\,\ell}(\mu_\mathrm{EW}) &=- \dfrac{v^2}{2\Lambda^2} \dfrac{1}{V_{ij}} \Big{[}C_{lequ}^{(3)}\Big{]}^\ast_{\ell \ell j i}\,.\nonumber
\end{align}

\noindent where we kept only the quark-flavor indices. From these relations, we see that contributions to $g_{V_R}^{ij}$ are necessarily lepton-flavor universal at dimension-6. Furthermore, the operators listed above also induce contributions to the di-lepton transitions $d_i\to d_j \ell\ell$, $d_i\to d_j \nu\nu$, $u_i\to u_j \ell\ell$ and $u_i\to u_j \nu\nu$. 

\paragraph{Operator mixing} Renormalization group equations (RGEs) are fundamental in order to relate the different scales involved in this problem. First, the running of the semileptonic operators from $\mu \approx 1~\mathrm{TeV}$ down to $\mu_\mathrm{EW}\approx m_W$ due to gauge interactions is given by~\cite{Gonzalez-Alonso:2017iyc}
\begin{equation}
\label{eq:rge-1}
\begin{pmatrix}
C_{lq}^{(3)}\\[0.35em] 
C_{ledq}\\[0.35em] 
C_{lequ}^{(1)}\\[0.35em] 
C_{lequ}^{(3)}
\end{pmatrix}_{(\mu=m_W)}
\approx\begin{pmatrix}
1.00 & 0 & 0 &0 \\[0.35em]
0 & 1.20 & 0 & 0\\[0.35em] 
0 & 0 & 1.20 & -0.19\\[0.35em]  
0& 0 & 0 & 0.96
\end{pmatrix}
\begin{pmatrix}
C_{lq}^{(3)}\\[0.35em] 
C_{ledq}\\[0.35em] 
C_{lequ}^{(1)}\\[0.35em] 
C_{lequ}^{(3)}
\end{pmatrix}_{(\mu=1~\mathrm{TeV})}\,,
\end{equation}

\noindent where we have omitted flavor indices and neglected the LFU operators. The $SU(3)_c\times U(1)_\mathrm{em}$ running below the EW scale reads~\cite{Gonzalez-Alonso:2017iyc}
\begin{equation}
\label{eq:rge-2}
\begin{pmatrix}
g_{V_L}\\[0.35em] 
g_{V_R}\\[0.35em] 
g_{S_L}\\[0.35em]
g_{S_R}\\[0.3em]
g_T
\end{pmatrix}_{(\mu=m_b)}
\approx\begin{pmatrix}
1.00  & 0 & 0 & 0 & 0 \\[0.35em]  
0 & 1.00 & 0 & 0 & 0\\[0.35em]  
0 & 0 & 1.46 & 0 & -0.02 \\[0.35em] 
0 & 0 & 0 & 1.46 & 0\\[0.35em] 
0 & 0 & 0 & 0 & 0.88
\end{pmatrix}
\begin{pmatrix}
g_{V_L}\\[0.35em] 
g_{V_R}\\[0.35em] 
g_{S_L}\\[0.35em]
g_{S_R}\\[0.35em]
g_T
\end{pmatrix}_{(\mu=m_W)}\,,
\end{equation}

\begin{table}[!t]
\renewcommand{\arraystretch}{2.0}
\centering
\begin{tabular}{|c|c|c|c|}
\hline 
SMEFT &  Definition & LEFT & LFU?\\ \hline\hline
$\Big{[}O_{lq}^{(3)}\Big{]}_{prst}$ &$\big{(}\bar{l}_p \gamma_\mu \tau^I l_r \big{)}\big{(}\bar{q}_s \gamma^\mu \tau^I  q_t \big{)}$ &  $g_{V_L}$ & \xmark\\[0.3em] 
$\Big{[}O_{ledq}\Big{]}_{prst}$ & $\big{(}\bar{l}_p^j e_r \big{)}\big{(}\bar{d}_s  q_t \big{)}+\mathrm{h.c.}$ &  $g_{S_R}$ & \xmark\\[0.3em]  
$\Big{[}O_{lequ}^{(1)}\Big{]}_{prst}$ &$\big{(}\bar{l}_p^j e_r \big{)}\epsilon_{jk}\big{(}\bar{q}_s^k u_t \big{)}+\mathrm{h.c.}$  &  $g_{S_L}$ & \xmark\\[0.3em] 
$\Big{[}O_{lequ}^{(3)}\Big{]}_{prst}$ & $\big{(}\bar{l}_p^j \sigma_{\mu\nu} e_r \big{)}\epsilon_{jk}\big{(}\bar{q}_s^k \sigma^{\mu\nu} u_t \big{)}+\mathrm{h.c.}$  &  $g_{T}$ & \xmark \\[0.3em]  \hline
$\Big{[}O_{Hl}^{(3)}\Big{]}_{pr}$ &$\big{(}H^\dagger i \overleftrightarrow{D_\mu} \tau^I H\big{)}\big{(}\bar{l}_p \gamma^\mu \tau^I l_r \big{)}$ & $g_{V_L}$ & \xmark\\[0.3em] 
$\Big{[}O_{Hq}^{(3)}\Big{]}_{pr}$ &$\big{(}H^\dagger i \overleftrightarrow{D_\mu} \tau^I H\big{)}\big{(}\bar{q}_p \gamma^\mu \tau^I q_r \big{)}$ & $g_{V_L}$ & \cmark\\[0.3em]  
$\Big{[}O_{Hud}^{(3)}\Big{]}_{pr}$ & $\big{(}\widetilde{H}^\dagger i {D_\mu} H\big{)}\big{(}\bar{u}_p \gamma^\mu  d_r \big{)}+\mathrm{h.c.}$ & $g_{V_R}$ & \cmark\\   \hline
\end{tabular}
\caption{ \sl \small SMEFT operators contributing to the low-energy EFT defined in Eq.~\eqref{eq:left}. Flavor indices are denoted by $\lbrace p,r,s,t\rbrace$ and $SU(2)_L$ indices by $\lbrace j,k\rbrace$. The operators $O_{Hq}^{(3)}$ and $O_{Hud}^{(3)}$ induce lepton-flavor universal (LFU) contributions. We use the same conventions of Ref.~\cite{Jenkins:2013zja}.}
\label{tab:smeft} 
\end{table}

\noindent and
\begin{equation}
\label{eq:rge-3}
 \begin{pmatrix}
g_{V_L}\\[0.35em] 
g_{V_R}\\[0.35em] 
g_{S_L}\\[0.35em]
g_{S_R}\\[0.35em]
g_T
\end{pmatrix}_{(\mu=2\,\mathrm{GeV})}
\approx\begin{pmatrix}
1.00 & 0 & 0 & 0 & 0\\[0.35em]  
0 & 1.00 & 0 & 0 & 0 \\[0.35em]  
0 & 0 & 1.72 & 0 & -0.02 \\[0.35em] 
0 & 0 & 0 & 1.72 & 0\\[0.35em] 
0 & 0 & 0 & 0 & 0.82
\end{pmatrix}
\begin{pmatrix}
g_{V_L}\\[0.35em] 
g_{V_R}\\[0.35em] 
g_{S_L}\\[0.35em]
g_{S_R}\\[0.35em]
g_T
\end{pmatrix}_{(\mu=m_W)} \,.
\end{equation}

\noindent In addition to these RGE effects, there are also the ones induced by the top-quark Yukawa, which mix the four-fermion operators with third-generation couplings into purely leptonic operators such as the ones contributing to $Z\to\ell\ell$~\cite{Feruglio:2016gvd} and $H\to\ell\ell$~\cite{Feruglio:2018fxo} which are of phenomenogical relevance. In summary, the combination of the tree-level matching relations in Eq.~\eqref{eq:smeft-matching}, with the RGE effects in Eq.~\eqref{eq:rge-1}--\eqref{eq:rge-3}, allows us to apply the constraints derived in these paper to any concrete NP scenario.

\section{Angular conventions}
\label{app:conventions}


\paragraph{Kinematics} Our conventions for the decay $P(p)\to P^\prime(k) \ell (k_1) \bar{\nu} (k_2)$ are summarized in Fig.~\ref{fig:semileptonic}. In the $P$ rest-frame, the leptonic and hadronic four-vectors $q=p-k$ and $k$ are given by
\begin{align}
 q^\mu= (q_0,0,0,q_z)\,,\qquad\qquad\quad k^\mu = (q_0,0,0,-q_z)\,,
\end{align}
where 
\begin{align}
\label{eq:semilep-kinematics}
 q_0=\frac{M^2+q^2-m^2}{2M}\,,\qquad k_0=\frac{M^2-q^2+m^2}{2M}\,,\quad\text{and}\quad q_z =\dfrac{\lambda^{1/2}(M^2,q^2,m^2)}{2M}\,.
\end{align}

\noindent In the dilepton rest-frame, the leptonic four-vectors read
\begin{align}
 k_1^\mu= (E_\ell, |p_\ell|\sin\theta_\ell,0,|p_\ell| \cos\theta)\,,\qquad\quad\quad k_2^\mu = (E_\nu, -|p_\ell|\sin\theta_\ell,0,-|p_\ell| \cos\theta)\,,
\end{align}
\noindent where
\begin{align}
\label{eq:semilep-kinematics-lep}
 E_\ell=\frac{q^2+m_\ell^2}{2\sqrt{q^2}}\,,
\end{align}
and $E_\nu=|p_\ell|=\sqrt{q^2}-E_\ell$.

\paragraph{Polarization vectors} In the $P$-meson rest-frame, we choose the polarization vectors of the virtual boson $V$ to be
\begin{align}
 \varepsilon^\mu(\pm) &= \frac{1}{\sqrt{2}}(0,\pm 1, i ,0)\,,\\[0.45em]
 \varepsilon^\mu(0) &= \frac{1}{\sqrt{q^2}}(q_z, 0, 0 , q_0)\,,\\[0.45em]
 \varepsilon^\mu(t) &= \frac{1}{\sqrt{q^2}}(q_0, 0, 0 , q_z)\,,
\end{align}
where $q_0$ and $q_z$ are given in Eq.~\eqref{eq:semilep-kinematics}. These four-vectors are orthonormal and satisfy the completeness relation~\eqref{eq:pol-completeness}.

\section{Form factor inputs}

The inputs needed to reproduce the form factor used in this paper are collected in Table~\ref{tab:references}.

\label{app:FFinputs}
\begin{table}[H]
\renewcommand{\arraystretch}{2.0}
\centering
\begin{tabular}{|c|c|c|c|c|}
\hline
\multicolumn{2}{|c|}{Form factor}                               & Ref.                                 & Parameterization                    & Numerical inputs \\ \hline\hline
\multirow{2}{*}{$K\to \pi$}              & S-V                  & \cite{Carrasco:2016kpy}    & $q^2$ expansion, Eq.~(53)                & Eq.~(54-61)         \\ 
                                         & T                    & \cite{Baum:2011rm}         & Simple pole, Eq.~(4)                  & Eq.~(11-12)         \\ \hline
\multirow{2}{*}{$D\to\pi$}               & S-V                  & \cite{Lubicz:2017syv}      & BGL, Eq.~(68-69)                      & Table 6-7        \\ 
                                         & T                    & \cite{Lubicz:2018rfs}      & BGL, Eq.~(31)                         & Table 7-8        \\ \hline
\multirow{2}{*}{$D\to K$}                & S-V                  & \cite{Lubicz:2017syv}      & BGL, Eq.~(70-71)                      & Table 8-9        \\ 
                                         & T                    & \cite{Lubicz:2018rfs}      & BGL, Eq.~(32)                         & Table 9-10       \\ \hline
\multirow{4}{*}{$B_{(s)}\to D_{(s)}$} & \multirow{2}{*}{S-V} & \cite{Lattice:2015rga}     & BGL, Eq.~(5.1-5.2)                       & Table IX         \\ 
                                         &                      & \cite{Na:2015kha}          & BCL, Eq.~(27-29, A1-A6)                  & Table VII        \\ 
                                         & \multirow{2}{*}{T}   & \cite{Atoui:2013zza}       & Ratio near zero recoil, Eq.~(37)      & Eq.~(52)            \\ 
                                         &                      & \cite{Bernlochner:2017jka} & z expansion, Eq.~(30-33) & Table II, IX     \\ \hline
\multirow{2}{*}{$B\to \pi$}              & S-V                  & \cite{Aoki:2019cca}        & BCL, Eq.~(448-449)                    & Table 41, 50         \\ 
                                         & T                    & \cite{Bailey:2015nbd}      & BCL, Eq.~(2-3)                        & Table II         \\ \hline
$B_s\to K$                               & S-V                  & \cite{Bazavov:2019aom}     & BCL, Eq.~(6.3-6.7b)                   & Table VIII-X     \\ \hline
\end{tabular}
\caption{Summary of parameterization and numerical inputs needed to compute scalar (S), vector (V) and tensor (T) form factors for each transition.}
\label{tab:references} 
\end{table}

\end{document}